\documentclass[onecolumn,showpacs,preprintnumbers,english,amsmath,amssymb]{revtex4}
\usepackage{graphicx,epsf,here}
\usepackage{dcolumn}
\usepackage{bm}

\newcommand{\Figurebb}[9]{
\begin{figure}[H]\begin{center}
\leavevmode
\epsfysize=#7cm
\epsfbox[#2 #3 #4 #5]{#6}
\par
\parbox{#8cm}{
\caption[figure]{\renewcommand{\baselinestretch}{0.8} \small
                                           \hspace{-0.3truecm}#9}
\label{#1}}
\end{center}
\end{figure}
}
\newcommand{\Table}[4]{
\begin{table}[H]\begin{center}{#3}
\parbox{#2cm}{
\vspace{0.5cm}
\caption[table]{\renewcommand{\baselinestretch}{0.8} \small
                \hspace{-0.3truecm}#4}
\label{#1}}
\end{center}
\end{table}
}
\def\fig#1{Fig.~\ref{#1}}
\def\tab#1{Table \ref{#1}}
\def\hom{\hbar\omega}

\def\siml{\,\hbox{\kern.1em \lower.6ex \hbox{$\sim$} \kern-1.12em
          \raise.6ex \hbox{$<$} }}
\def\simg{\,\hbox{\kern.1em \lower.6ex \hbox{$\sim$} \kern-1.12em
          \raise.6ex \hbox{$>$} }}
\def\eq#1{(\ref{#1})}
\def\papa#1#2{\frac{\partial#1}{\partial#2}}
\def\etal{{\it et al.}}
\def\bfr{{\bf r}}
\def\bfp{{\bf p}}
\def\bfq{{\bf q}}
\def\bea{\begin{eqnarray}}
\def\eea{\end{eqnarray}}
\def\bgrk#1{\mbox{{\boldmath $#1$ \unboldmath}}\!\!}

\begin{document}

\title{Uniform semiclassical trace formula for U(3) $\to$ SO(3) 
       symmetry breaking}
\date{October 4, 2005}
\author{M. Brack}
\affiliation{Institut f\"{u}r Theoretische Physik, Universit\"{a}t
Regensburg, D-93040 Regensburg, Germany}
\author{M. \"{O}gren}
\author{Y. Yu}
\author{S. M. Reimann}
\affiliation{Division of Mathematical Physics, LTH, Lund University. P.O. Box
118, S-221 00 Lund, Sweden}
\pacs{05.30.Fk}

\begin{abstract}
We develop a uniform semiclassical trace formula for the density of states 
of a three-dimensional isotropic harmonic oscillator (HO), perturbed by a term 
$\frac14 \epsilon\,r^4$. This term breaks the U(3) symmetry of the HO, 
resulting in a spherical system with SO(3) symmetry. We first treat the 
anharmonic term for small $\epsilon$ in semiclassical perturbation theory by 
integration of the action of the perturbed periodic HO orbit families over 
the manifold $\mathbb{C}$P$^2$ which is covered by the parameters 
describing their four-fold degeneracy. Then we obtain an analytical uniform 
trace formula for arbitrary $\epsilon$ which in the limit of strong 
perturbations (or high energy) asymptotically goes over into the correct 
trace formula of the full anharmonic system with SO(3) symmetry, and in the 
limit $\epsilon$ (or energy) $\to 0$ restores the HO trace formula with U(3) 
symmetry. We demonstrate that the gross-shell 
structure of this anharmonically perturbed system is dominated by the two-fold 
degenerate diameter and circular orbits, and {\it not} by the orbits with the 
largest classical degeneracy, which are the three-fold degenerate tori with 
rational ratios $\omega_r:\omega_\varphi=N\!:\!M$ of radial and angular 
frequencies. The same holds also for the limit of a purely quartic spherical
potential $V(r)\propto r^4$.

\end{abstract}

\maketitle 

\section{Introduction}

The semiclassical quantisation of non-integrable systems using properties
of their periodic orbits was triggered by M. Gutzwiller \cite{gutz} and
extended by several groups \cite{bablo,struma,crli1,crli2}. It allows one 
to express the oscillating part $\delta g(E)$ of the quantum-mechanical 
density of states, given exactly in terms of a quantum spectrum $E_n$ and
separated into two terms by
\bea
g(E) = \sum_n \delta(E-E_n) = {\widetilde g}\,(E)+\delta g(E)\,,
\label{gqm}
\eea
through a semiclassical trace formula of the form
\bea
\delta g_{sc}(E) \simeq \sum_{po} {\cal A}_{po}(E) 
                        \cos\,[S_{po}(E)/\hbar-\sigma_{po}\pi/2]\,.
\label{dgsc}
\eea
The sum is over all periodic orbits $(po)$ of the classical system, 
$S_{po}(E)=\oint\bfp\cdot d\bfq$ are their action integrals, the amplitudes
${\cal A}_{po}(E)$ depend on their stabilities and degeneracies, and
$\sigma_{po}$ are some phases called Maslov indices. The average part
${\widetilde g}\,(E)$ of the density of states, which by definition varies
smoothly with energy, is obtained by the extended Thomas-Fermi (ETF) model 
(see, e.g., \cite{book}, Chapter 4). Then, the sum $g_{ETF}(E)+\delta g_{sc}(E)$ 
usually turns out to be a good approximation to the exact quantity \eq{gqm},
although the sum over the $po$ in \eq{dgsc} is only an asymptotic one,
correct to leading order in $1/\hbar$, and in chaotic systems is hampered by 
convergence problems \cite{chaos}. 
For integrable systems, Berry and Tabor \cite{bertab} and later Creagh \etal\ 
\cite{crli2} showed how a trace formula of the form \eq{dgsc} can quite 
generally be obtained from the Einstein-Brillouin-Keller (EBK) quantisation 
\cite{ebk}. This method will be used in the present paper for a spherical
system.

The semiclassical theory, often referred to as periodic orbit theory (POT), 
has been very successful not only in (partial) quantisation of a given
Hamiltonian, but also in interpreting many experimentally observable quantum 
oscillations in finite fermion systems -- so-called ``shell effects'' -- in 
terms of classical mechanics. Examples are atomic nuclei \cite{bm}, metallic 
clusters \cite{nish,klavs,mbclus}, semiconductor quantum dots \cite{steffi}, 
and metallic nanowires \cite{wire}. An overview over many aspects of the POT 
and illustrative applications are given in \cite{book}.

One problem that remains with all the trace formulae developed in the work 
quoted above is that the amplitudes ${\cal A}_{po}$ diverge in situations 
where a continuous symmetry is broken or restored under the variation of a 
system parameter (which may also be the energy $E$), or where bifurcations 
of periodic orbits occur. In such situations one has to go beyond the 
stationary phase approximation which is underlying the semiclassical approach. 
This has, besides \cite{bablo,struma,crli1,crli2,bertab}, been developed most 
systematically by Ozorio de Almeida and Hannay \cite{ozoha} for both 
situations, leading to local uniform approximations with finite amplitudes
${\cal A}_{po}$. So-called global uniform approximations, which yield finite 
amplitudes at symmetry-breaking and bifurcation points and far from them go 
over into the standard (extended) Gutzwiller trace formula, were developed 
for the breaking of U(1) symmetry in \cite{toms}, for some cases of U(2) and 
SO(3) symmetry breaking in \cite{hhuni}, and for various types of bifurcations 
in \cite{ssun}. (Details and further references may be found in \cite{book}, 
Chapter 6.3.)

In this paper we investigate an anharmonically perturbed three-dimensional 
spherical harmonic oscillator (HO) Hamiltonian (for a particle with mass $m=1$):
\bea
H(\bfr,\bfp) = \frac12\,p^2+\frac12\,\omega^2 r^2+\frac14\,\epsilon\,r^4
             = H_0(r,p) + \epsilon\,\delta H(r)\,,
\label{hpert}
\eea
where $p=|\bfp|$ and $r=|\bfr|$ are the absolute values of the 
three-dimensional momentum and radius vectors. $\epsilon>0$ is the strength 
of the perturbation which first is assumed to be small, but later may 
assume arbitrary positive values. The unperturbed HO system has U(3) symmetry 
\cite{lipkin} which is broken by the anharmonic term, leading to the SO(3)
symmetry of the perturbed spherical system. 

The Hamiltonian \eq{hpert} suggested itself in a recent study \cite{yylund} 
of harmonically trapped fermionic atoms with a short-range repulsive two-body
interaction, treated self-consistently in the Hartree-Fock approximation. 
As a result, very pronounced shell effects in the total energy $E_{tot}$ of 
the interacting system as a function of the number $N$ of atoms were found, 
which remind about the so-called super-shells predicted \cite{nish} and 
observed \cite{klavs} in metallic clusters. In a first attempt to interpret 
these shell effects semiclassically \cite{yylund}, we parameterised the 
self-consistent mean field of the interacting system by the Hamiltonian 
\eq{hpert} and applied the semiclassical perturbation theory of Creagh 
\cite{crpert} to explain qualitatively the shell structure of the HF results.

In the present paper we describe some of the mathematical details of the 
semiclassical perturbation theory for the symmetry breaking U(3) $\to$ 
SO(3) and develop a uniform trace formula valid for arbitrary strengths 
$\epsilon$ in the Hamiltonian \eq{hpert}. In Section \ref{secpert} we present 
the perturbative trace formula for the density of states which is valid in the 
limit of small $\epsilon$ and has been presented shortly in \cite{yylund}. It 
already puts the dominance of the shortest periodic orbits, namely the circles 
and diameters, into evidence. Although their individual contributions to the 
perturbative trace formula diverge in the limit $\epsilon\to 0$, their 
sum restores to the unperturbed HO trace formula with U(3) symmetry in this 
limit, which at the same time is the limit of zero energy.

In Section \ref{secuni} we develop the uniform trace formula that includes 
the contributions of the diameter and circle orbits for arbitrary values of 
$\epsilon$. For this purpose, we start from the EBK quantisation of an 
arbitrary system with spherical symmetry and apply the Poisson summation 
formula. We obtain a one-dimensional trace integral for the semiclassical 
density of states, whose end-point contributions correspond to the diameter 
and circle orbits, valid for an arbitrary spherical one-body potential.
For the Hamiltonian \eq{hpert}, the gross-shell structure of the density of 
states is at low energies always dominated by the families of diameter and 
circle orbits, although these orbit families only have a two-fold degeneracy.
At sufficiently high energies and repetition numbers, ``rational tori'' with 
frequency ratios $\omega_r\!:\!\omega_\phi=N\!:\!M>2$ with $N\geq 7$ bifurcate from 
the circle orbits with repetition numbers $M\geq 3$, as discussed in Section 
\ref{sectori}. These rational tori have the highest (three-fold) degeneracy 
possible in a three-dimensional spherical system, but due to their length 
they only affect the finer quantum structures of the density of states. This
situation is completely different from that of a spherical billiard \cite{bablo}
where the contribution of the diameter orbits to the density of states is
practically negligible. 

In the limit $\epsilon\to\infty$, \eq{hpert} becomes a purely quartic 
oscillator which in many respects is easier to handle. In particular, in this 
limit the system acquires the ``scaling property'' that its classical dynamics 
becomes independent of the energy which can be absorbed by a rescaling of 
coordinates and time. This limit will be discussed separately in Section 
\ref{secquart}. One interesting classical aspect is that no bifurcations of 
the circle orbits occur. The same rational tori, which bifurcate from the 
circle orbit in the perturbed HO system, exist here at all energies. But, 
again, they affect only the finer quantum structures of the density of states, 
while its gross-shell structure is largely dominated by the circles and 
diameters.

In the appendices \ref{app1} - \ref{app3}, we collect some mathematical details 
about the integration over the manifolds $\mathbb{C}$P$^2$ and S$^5$ and some 
explicit analytical expressions for action integrals and periods in terms of 
elliptic integrals. Finally, in the appendix \ref{appothers}, we re-derive 
from our general trace integral the known trace formulae for two popular 
spherical systems: the spherical billiard and the Coulomb potential.
 
\newpage

\section{Perturbative treatment of the anharmonicity}
\label{secpert}

We first write down the exact density of states of the unperturbed 
three-dimensional harmonic oscillator (HO), given by the Hamiltonian
\bea
H_0(r,p) = \frac12\, p^2 + \frac12\,\omega^2 r^2\,.
\label{hho}
\eea
Its quantum-mechanical density of states can be written in the form
\bea
g_0(E) = \sum_{n=0}^\infty d_n \delta(E-E_n)
       = g_0^{ETF}(E) + \delta g_0(E)\,,
\label{gsum}
\eea
using the spectrum $E_n = \hom(n+3/2)$ with the degeneracy factor
$d_n = (n+1)(n+2)/2$. Here $g_0^{ETF}(E)$ is the smooth part given by the 
ETF model
\bea
g_0^{ETF}(E) = \frac{1}{2(\hom)^3} \left[ E^2 - \frac14 (\hom)^2 \right],
\label{getfho}
\eea
and $\delta g_0(E)$ is the oscillating part given by the following exact
trace formula \cite{book}
\bea
\delta g_0(E) = 2\,g_0^{ETF}(E)\sum_{k=1}^\infty (-1)^k
                \cos\left(2\pi k\frac{E}{\hom}\right)
              = 2\,g_0^{ETF}(E)\sum_{k=1}^\infty (-1)^k
                \cos[kS_0(E)/\hbar]\,.
\label{dg3ho}
\eea
It can be understood as a sum over all periodic orbits of the system. Hereby 
$k$ represents the repetition number of the primitive classical periodic 
orbit which has the action $S_0(E)=2\pi E/\omega$. Keeping only the leading TF 
term of $g_0^{ETF}(E)$ in \eq{getfho}, the density of states can also be 
written in the form
\bea
g_0(E) = \frac{E^2}{2(\hom)^3}\;
         \Re e \!\sum_{-\infty}^\infty (-1)^k\,e^{i2\pi kE/\hom} 
       + {\cal O}(\hbar^{-1})\,,
\label{g3hoc}
\eea
since the imaginary parts cancel upon summation. Note that the smooth TF part 
in \eq{getfho} comes from the contribution with $k=0$ in \eq{g3hoc}.

Next we follow Creagh \cite{crpert} by writing the perturbed trace
formula in the form
\bea
g_{pert}(E) = \frac{E^2}{2(\hom)^3}\;\Re e \!\sum_{-\infty}^\infty (-1)^k\,
              {\cal M}(k\sigma/\hbar)\,e^{i2\pi kE/\hom},
\label{gpert}
\eea
where ${\cal M}(k\sigma/\hbar)$ is a modulation factor that takes into 
account the lowest-order perturbation of the action of the HO orbits. 
Here $\sigma=\sigma(\epsilon,E)$ is a small action that, quite generally, 
depends on $\epsilon$ and the energy $E$. The factor in front of the sum
in \eq{gpert} takes into account only the leading term of the smooth part 
$g_0^{ETF}(E)$; this is consistent with the fact that the perturbation 
theory only deals with the terms of leading order in $\hbar^{-1}$ of the 
semiclassical trace formula. As we will see in Section \ref{secqm}, the
contributions of terms of relative order $\hbar^2$ in $g_0^{ETF}(E)$ are,
indeed, negligible.

As described in detail in \cite{crpert}, one obtains 
${\cal M}(k\sigma/\hbar)$ from an integration of a perturbative phase 
function over the manifold that describes the classical degeneracy of the 
HO orbits. The unperturbed HO has U(3) symmetry, so that 
${\cal M}(k\sigma/\hbar)$ is formally obtained by the average 
\bea
{\cal M}(k\sigma/\hbar) = \langle e^{ik\Delta S(o)/\hbar}\rangle_{o\in U(3)},
\label{mav}
\eea  
where $o$ is an element of the group U(3) characterising a member of 
the unperturbed HO orbit family, and $\Delta S(o)$ is the lowest-order
primitive action shift brought about by the perturbation $\epsilon\,\delta 
H(r)$ in \eq{hpert}. In the present case $\Delta S(o)$ is nonzero already 
in first-order perturbation theory and therefore $\sigma$ is proportional 
to $\epsilon$.

We need, however, not integrate over the full nine-dimensional space of 
group elements of U(3). It is sufficient to consider only the subset of
symmetry operations which transform the orbits within a given degenerate 
family into each other (without changing their actions). The dimension of
this subset is the degree of degeneracy $f$ of that family. For the 
three-dimensional spherical HO we have $f=4$. This can most easily be seen 
by the following argument which also will allow us to find a suitable
parametrisation for the four-fold integration. The full phase space 
is six-dimensional; due to energy conservation it is reduced to a 
five-dimensional energy shell which has the topology of a five-sphere S$^5$. 
Of the five parameters that specify a point on S$^5$, one can be chosen as 
a trivial time shift along the periodic orbits corresponding to a simple 
phase factor $e^{-i\omega t}$. This parameter forms a subgroup U(1), so 
that its elimination restricts us to the four-dimensional manifold
(see Appendix \ref{app1} for its mathematical definition)
\bea
S^5\!/U(1) = \mathbb{C}{\rm P}^2,
\label{coset}
\eea 
which is neither 
a four-torus nor a four-sphere \cite{bbz}. (Note that for the two-dimensional HO 
one is led to the manifold $\mathbb{C}$P$^1$ which happens to be homomorphous 
to the two-sphere S$^2$, see \cite{crpert}.) A suitable parametrisation 
of $\mathbb{C}$P$^2$ is given in Appendix \ref{app1} in terms of four angles 
$(\vartheta,\varphi,\nu_2,\nu_3)$ whose meaning will become clear in a moment. 
The modulation factor therefore becomes
\bea
{\cal M}(k\sigma/\hbar) = \frac{2}{\pi^2}\!\int d\Omega_{\mathbb{C}{\rm P}^2}\,
                          e^{ik\Delta S/\hbar}
                        = \frac{2}{\pi^2}\!\int_0^{\frac{\pi}{2}}\! 
                         \cos\varphi\sin\varphi\,d\varphi
                         \!\int_0^{\frac{\pi}{2}}\! 
                         \sin^3\!\vartheta\cos\vartheta\,d\vartheta
                         \!\int_0^{2\pi}\! d\nu_2 \!\int_0^{2\pi}\! d\nu_3\,
                         e^{ik\Delta S(\vartheta,\varphi,\nu_2,\nu_3)/\hbar}.
\label{mfac}
\eea
In the zero-perturbation limit $\epsilon\to 0$, where the action 
shift $\Delta S$ and hence also $\sigma$ is zero, the modulation factor 
becomes unity, as it should for \eq{gpert} to approach \eq{g3hoc}.

Next we have to determine the action shift $\Delta S$ of a primitive
periodic orbit of the HO, caused by the perturbation $\epsilon r^4/4$.
The harmonic solutions of the classical equations of motion of the
unperturbed system are well known:
 \bea
x_i = \sqrt{2E_i/\omega^2}\cos(\omega t+\nu_i)\,,
      \qquad i=1,2,3 \qquad (x_1,x_2,x_3)=(x,y,z).
\label{xi}
\eea
Hereby $E_i$ are the three conserved energies in the three dimensions. 
Depending on the values of $E_i$ and the phases $\nu_i$, \eq{xi}
describes circles, ellipses or librations through the origin; we will
in the following call the latter the ``diameter orbits''. We now
re-parameterise the $x_i$ in the following way:
\bea
x(t) = R\,n_1\,\cos(\omega t)\,,\qquad
y(t) = R\,n_2\,\cos(\omega t+\nu_2)\,,\qquad
z(t) = R\,n_3\,\cos(\omega t+\nu_3)\,,
\label{xi3d}
\eea
where $\nu_2,\nu_3\in[0,2\pi)$, and $R$ is given by
\bea
R=\sqrt{2E}/\omega\,,\qquad E=E_1+E_2+E_3\,.
\eea
Since the $(n_1,n_2,n_3)$ must lie on the sphere $S^2$, due to the
conservation of energy, we can also write them as:
\bea
n_1 = \cos\vartheta\,,\qquad
n_2 = \sin\vartheta\,\cos\varphi\,,\qquad
n_3 = \sin\vartheta\,\sin\varphi\,.
\label{s2par}
\eea
Actually, since the $n_i$ must be restricted to positive definite
values in order to cover once all classically allowed values of the 
$x_i(t)$, we only have to integrate over the first octant of S$^2$,
so that $\varphi,\vartheta\in[0,\pi/2]$. The four angles 
$\vartheta,\varphi,\nu_2,\nu_3$, with their ranges of definition, are 
precisely the parameters describing the manifold $\mathbb{C}$P$^2$ 
as explained in Appendix \ref{app1}, and the correct integration measure 
is that used in \eq{mfac}. We might have kept the phase angle $\nu_1$ 
of $x(t)$ in \eq{xi3d}, too; integration over it is tantamount to a 
trivial integration over time along the orbits. The resulting 
five-dimensional integral becomes equivalent to that over the S$^5$ 
sphere discussed in Appendix \ref{app2}.

According to Creagh \cite{crpert}, the first-order action shift
brought about by a perturbation $\epsilon\delta H$ is given by
\bea
\Delta_1 S = - \epsilon \oint_{po} \delta H(t) dt\,,
\eea
where $po$ stands for a particular member of the unperturbed periodic orbit 
family, and the perturbating Hamiltonian $\delta H(t)=\delta H(x_i(t))$ has 
to be evaluated along the periodic orbit. Inserting \eq{xi3d} - \eq{s2par} 
into the perturbation $\delta H = r^4\!/4 = (x^2+y^2+z^2)^2\!/4$ leads to 
elementary integrals over powers of the trigonometric functions. The result is
\bea
\Delta_1 S(\vartheta,\varphi,\nu_2,\nu_3)
       & = & -\epsilon\frac{\pi}{16\,\omega}\,(3R^4-4\,L^2\!/\omega^2)
       \;=\; -\frac{\epsilon\pi}{4\,\omega^5}\,(3E^2-\omega^2L^2)\nonumber\\
       & = & -\,\sigma\,[\,3-\sin^2(2\vartheta)\,
             (\cos^2\!\varphi\,\sin^2\!\nu_2+\sin^2\!\varphi\,\sin^2\!\nu_3)
             -\sin^4\!\vartheta\,\sin^2(2\varphi)\,\sin^2(\nu_2-\nu_3)]\,,
\label{ds3d}
\eea
where $L^2=L_x^2+L_y^2+L_z^2$ is the conserved squared angular momentum,
and the first-order action unit $\sigma$ is given by 
\bea
\sigma = \epsilon\pi E^2\!/4\,\omega^5.
\label{sigma}
\eea

The integral \eq{mfac}, with the function given in \eq{ds3d}, has been
integrated numerically and found to be identical with the corresponding
five-dimensional integral over the 5-sphere S$^5$, expressing $\Delta _1S$
in terms of the five hyperspherical angles of the six-dimensional polar
coordinates (cf.\ Appendix \ref{app2}). However, even the
four-dimensional integration in \eq{mfac} is more than needed. In fact, 
the same result is obtained if we replace the phase function under the 
$\mathbb{C}$P$^2$ integration \eq{mfac} as follows:
\bea
\Delta_1 S(\vartheta,\varphi,\nu_2,\nu_3) \quad \Longrightarrow \quad
\Delta_1 S(\varphi) \;=\; -\,\sigma\,(3-\cos^2\!\varphi).
\label{ds3da}
\eea
The integrals over $\vartheta$, $\nu_2$ and $\nu_3$ then become trivial and 
the result is found (with $z=\cos^2\varphi$) to be
\bea
{\cal M}(k\sigma/\hbar) & = & 2 \!\int_0^{\frac{\pi}{2}}\! 
                              \cos\varphi\sin\varphi\,d\varphi\,
                              e^{-ik\sigma(3-\cos^2\!\varphi)/\hbar}  
                          = \int_0^1 dz\,e^{-ik\sigma(3-z)/\hbar}
                          = \frac{\hbar}{k\sigma}\!
                            \left[\,e^{-i2k\sigma/\hbar-i\pi/2}
                            +e^{-i3k\sigma/\hbar+i\pi/2}\right]\!.~~~~~
\label{modfex}
\eea
The replacement \eq{ds3da} is not obviously justifiable in a direct way, but 
a suitable reduction of the five-dimensional S$^5$ integral yields exactly 
the integral \eq{modfex}, as demonstrated in Appendix \ref{app2}. Identically 
the same result is also obtained independently from the EBK quantisation of 
the perturbed Hamiltonian using Poisson summation and keeping only first-order
terms in $\epsilon$; it is given in Eq.\ \eq{gebkp} at the end of Sect.\ 
\ref{secpois}.

The form \eq{modfex} of the modulation factor suggests a simple physical 
interpretation: the resulting two terms correspond to two separate families 
of orbits that survive the breaking of the U(3) symmetry. As we will see below, 
these are the circle and the diameter orbits with maximal and minimal (i.e., 
zero) angular momentum, respectively, at fixed energy. For $k=1$ these are 
actually the shortest periodic orbits found in the perturbed system. In 
contrast to the four-fold degenerate unperturbed HO orbits, they only have a 
two-fold degeneracy, since only rotations about two of the Euler angles change
their orientation. In general, the most degenerate periodic orbits in a 
three-dimensional system with spherical symmetry SO(3) can be rotated about 
three Euler angles. However, for each of the circle and diameter orbits one of 
the three rotations is redundant in the sense that it maps these orbits onto
themselves: for the circles, it is the in-plane rotation around their centre, 
and for the diameters, it is the rotation about their own axis of motion. The 
loss of two degrees of degeneracy of these two orbit types with respect to the 
unperturbed HO orbits appears in the form of the factor $\hbar$ in their 
amplitudes in \eq{modfex}, in agreement with the general counting of the powers 
of $\hbar$ in semiclassical amplitudes \cite{book,crli1}. 

We will see in the next section that for values $k\geq 3$ and large enough 
values of $\epsilon$, there exist fully three-fold degenerate orbits, namely 
orbits with rational ratios $\omega_r:\omega_\varphi = N\!:\!M$ of radial and 
angular frequencies with $N\!:\!M\geq 7\!:\!3$. They are created at bifurcations 
of the $k$-the repetitions of the circle orbits with $k=|M|\geq 3$. This 
happens, however, only at finite values of $\epsilon$, so that these orbits 
do not contribute in the limit $\epsilon\to 0$.

Inserting \eq{modfex} into \eq{gpert}, we find the following perturbed
trace formula for the oscillating part of the level density:
\bea
\delta g_{pert}(E) & = & \frac{4\omega^2}{\epsilon\hbar^2\pi}
                         \sum_{k=1}^\infty \frac{(-1)^k}{k} 
                         \left[\cos\left(\frac{kS_c}{\hbar}-\frac{\pi}{2}\right)\!
                         +\cos\left(\frac{kS_d}{\hbar}-\frac{3\pi}{2}\right)\right]
                         \nonumber\\
                   & = & \frac{4\omega^2}{\epsilon\hbar^2\pi}
                         \sum_{k=1}^\infty \frac{(-1)^k}{k}\,
                         \left[\,\sin(kS_c/\hbar)-\sin(kS_d/\hbar)\,\right],
\label{dgp}
\eea
where $S_d$ and $S_c$ are the perturbed primitive actions of the diameters 
and the circle, respectively:
\bea
S_d=S_0-3\sigma\,, \qquad S_c=S_0-2\sigma\,,\qquad S_0=2\pi E/\omega\,.
\label{sorbits}
\eea
These values will be confirmed from the general expressions derived in the
next section. The perturbative trace formula \eq{dgp} uniformly restores
the oscillating part of the exact HO trace formula \eq{g3hoc} in the limit
$\epsilon\to 0$. In Section \ref{secqm} we will test its range of 
validity comparing against quantum-mechanical results.

Note that -- as is usual in perturbation theory -- the semiclassical
amplitudes and actions of the orbits contributing to \eq{dgp} are correct
only in the small-$\epsilon$ limit. In general, one has to generalise the
perturbative trace formula to a uniform version that, in the limit of
large perturbation, goes over into the corresponding (extended)
Gutzwiller trace formula. This uniformisation has been done for
the breaking of U(1) symmetry in \cite{toms}, and for some cases of
U(2) symmetry breaking in \cite{hhuni}; one of the latter result applies,
with suitable changes, also to some cases of SO(3) $\to$ U(1)
symmetry breaking. The symmetry breaking U(3) $\to$ SO(3) under
study here has not been treated in the literature so far. However, the
simplicity of our above results makes the uniformisation particularly 
easy in the present case: since the modulation factor \eq{modfex}
already has its own asymptotic form -- or, inversely speaking: since
the asymptotic expansion of the integral in \eq{modfex} in the limit
of large $\sigma$ happens to be exact also for $\sigma\to 0$
-- it will be sufficient to replace in \eq{dgp} the perturbed actions
$S_d$, $S_c$ and their semiclassical amplitudes by those valid for all 
values of $\epsilon$, which will be derived in the next section.

\newpage

\section{Uniform trace formula for diameter and circle orbits}
\label{secuni}

In this section we will calculate the full actions and semiclassical
amplitudes of the diameter and circle orbits of the Hamiltonian \eq{hpert},
valid for arbitrary values of $\epsilon$. The general trace formula for
an arbitrary spherical potential in three dimensions has been given by
Creagh and Littlejohn \cite{crli1,crli2}. We choose a different approach 
here, which is in spirit that of Berry and Tabor \cite{bertab}, but goes 
beyond their leading-order approximation in taking into account the 
end-point corrections to a trace integral which yield precisely the circle 
and diameter orbit contributions.

The classical Hamiltonian of the system (with mass $m=1$)
\bea
H(\bfr,\bfp) = E = \frac12\,p^2 + V(r)\,,\qquad 
        V(r) = \frac12\,\omega^2 r^2+\frac14\,\epsilon\, r^4\,,
\label{hrad}
\eea
is integrable due to the spherical symmetry of the potential, so that we 
can apply the standard EBK (or radial WKB) approximation to it. Writing $H$ 
in terms of polar coordinates $\bfr=(r,\theta,\phi)$ and the associated 
canonical momenta $\bfp = (p_r,p_\theta,p_\phi)$, we have the usual form 
involving an effective potential $V_{eff}(r)$ that includes a centrifugal 
term:
\bea
E \; = \; H(r,p_r,L) = \frac12\, p_r^2 + V_{eff}(r)\,, \qquad 
V_{eff}(r) = V(r) + \frac{L^2}{2r^2}\,,
\label{heff}
\eea
where $p_r$ is the radial momentum. The three independent (and 
Poisson-commuting) constants of the motion are the energy $E$, the total 
angular momentum $L^2$, and its $z$ component $L_z=p_\phi$. The momentum 
$p_\theta$ can be expressed in terms of the latter two as
\bea
p_\theta = \sqrt{L^2-L_z^2/\!\sin^2\theta}\,.
\label{pthet}
\eea
Before we specialise to our particular potential $V(r)$ in \eq{hrad}, we
briefly recall the radial EBK quantisation and derive from it a
general trace formula for an arbitrary spherical potential, starting
from the Hamiltonian \eq{heff}.

\subsection{EBK quantisation of a spherical system}
\label{secebk}

In the standard radial EBK method \cite{ebk,ajp}, one quantises the three 
following action integrals
\bea
S_r      & = & \oint p_r\,dr = 2\pi\hbar\,(n_r+1/2)\,,
               \;\,\qquad\qquad n_r = 0,1,2,\dots \label{Irad}\\
I_\theta & = & \frac{1}{2\pi}\oint p_\theta\,d\theta = \hbar\,(n_\theta+1/2)\,,
               \qquad\qquad n_\theta = 0,1,2,\dots \label{Ithet}\\
I_\phi   & = & L_z = \hbar\, m\,,\hspace*{4.05cm} 
               m=0,\pm 1,\pm 2,\dots\label{Iphi}  
\eea
where the Maslov index 1/2 in \eq{Irad} is correct only for smooth potentials.
Since $p_\theta$ in \eq{pthet} does not depend on the potential, the integral 
for $I_\theta$ can be done once for all and yields
\bea
I_\theta = L-|L_z| \geq 0\,,
\label{Ithet1}
\eea
so that the quantisation condition for $L$ is given by
\bea
L = \hbar\,(n_\theta+|m|+1/2) = \hbar\,(\ell+1/2)\,,\qquad\qquad
    \ell=n_\theta+|m|=0,1,2, \dots
\label{Lquant}
\eea
in terms of a single angular momentum quantum number $\ell$. The relation 
\eq{Lquant} includes the so-called Langer correction; the quantised squared 
angular momentum $L^2=\hbar^2\,(\ell+1/2)^2=\hbar^2\,(\ell^2+\ell+1/4)$ 
agrees with the exact quantum-mechanical value $\hbar^2\,\ell(\ell+1)$ in 
the limit of large $\ell$. Solving \eq{heff} for $p_r$, we can write the 
radial action as
\bea
S_r(E,L) = \oint dr\sqrt{2E-2V(r)-L^2\!/r^2} = 2\pi\hbar\,(n_r+1/2)\,,
\label{Irad1}
\eea
showing that the quantised energies will only depend on the radial and
angular momentum quantum numbers $n_r$ and $\ell$; they have, of course,
the usual $m$-degeneracy $d_\ell=(2\ell+1)$, since $-\ell\leq m\leq +\ell$.

Inverting the relation \eq{Irad1}, we may rewrite the Hamiltonian \eq{heff}
in the form
\bea
E = \widetilde{\!H}(S_r,L)\,.
\label{ham}
\eea
Inserting the right-hand side of \eq{Irad} and \eq{Lquant} into \eq{ham}, 
we obtain the EBK-quantised eigenenergies:
\bea
E_{n_r\ell}^{ebk} = \widetilde{\!H}(2\pi\hbar\,(n_r+1/2),\hbar\,(\ell+1/2))\,,
                    \qquad\qquad \quad \ell,n_r = 0,1,2,\dots
\label{eebk}
\eea
They can in general only be obtained by numerical iteration after doing 
the radial action integral \eq{Irad1} over $r$ within the classical turning 
points.

\subsection{Introduction of scaled variables}
\label{secscal}

Before continuing, we simplify the situation by a scaling of the energy. 
In principle we have to vary the three parameters $E$, $\omega$, and 
$\epsilon$ to study the dynamics of our present system. However, we can 
introduce a scaling of the energy in such a way that the classical dynamics 
only depends on one single parameter, a dimensionless scaled energy $e$. 
If we multiply \eq{hpert} by the factor $\epsilon/\omega^4$, we can write 
the r.h.s.\ in terms of scaled coordinates $q_i$ and momenta ${\tilde p}_i$ 
and a scaled time $\tau$:
\bea
         q_i=\frac{\sqrt{\epsilon}}{\omega^2}\,x_i\,,\qquad
{\tilde p}_i=\frac{\sqrt{\epsilon}}{\omega^3}\,p_i\,,\qquad
        \tau=\omega t\,,
\label{scal} 
\eea
so that the scaled energy $e$ becomes
\bea
e = \frac{\epsilon}{\,\omega^4}E = \frac12\,{\bf \tilde{p}}^2+v(q)
                                 = \frac12\,{\dot q}^2+v(q)+\frac{l^2}{2q^2}\,,
    \qquad v(q)=\frac12\,q^2+\frac14\,q^4\,.
\label{escal}
\eea
where the dimensionless scaled angular momentum $l$ is given by
\bea
l=L/s\,,\qquad s = \frac{\,\omega^3}{\epsilon}\,.
\label{sscal}
\eea
In \eq{escal} $q$ is the scaled radial variable and ${\dot q}=\tilde{p}_q$ the scaled 
radial momentum, and the dot means
the derivative with respect to the scaled time $\tau$. $s$ is the action unit.

This brings about a considerable simplification of the classical dynamics:
we only need to vary one parameter $e$; at the end of our calculations 
we just have to remember that energies are measured in units of 
$\omega^4\!/\epsilon$, angular momentum and actions in units of 
$s=\omega^3\!/\epsilon$, times in units of $1/\omega$, etc. In the following
we shall give all quantities as functions of the dimensionless scaled
variables $e$ and $l$. (Other scaled dimensionless quantities such
as actions, periods etc.\ will be denoted by lower-case letters.)

The scaling in \eq{escal} is specific for our present Hamiltonian \eq{hrad}.
It can, however, easily be modified for HO potentials perturbed by arbitrary 
central potentials which are pure power laws in $q$. The results of this
subsection can therefore easily be generalized to the corresponding suitably 
scaled potentials $v(q)$. 

\subsection{Density of states for an arbitrary shperical potential}
\label{secpois}

We now take the spectrum \eq{eebk} as a starting point to write down the
density of states in the EBK approximation:
\bea
g_{ebk}(e) = \sum_{n_r=0}^\infty \sum_{\ell=0}^\infty
             (2\ell+1)\,\delta\!\left(e-e_{n_r\ell}^{ebk}\right).
\label{gebk}
\eea
Next we apply Poisson summation \cite{titch} to convert the sums over $n_r$ 
and $\ell$ into integrals:
\bea
g_{ebk}(e) = \sum_{N=-\infty}^\infty \sum_{M=-\infty}^\infty
             \int_0^\infty d\ell\, (2\ell+1) \int_0^\infty dn_r\,
             \delta(e-e_{n_r\ell}^{ebk})\,
             e^{i2\pi(Nn_r+M\ell)} + \dots
\label{gebk1}
\eea
Due to the finite lower limits of the summations in \eq{gebk}, there
are boundary corrections to \eq{gebk1}, which we have indicated by the
dots. We shall comment on them in a moment. Using \eq{Irad} and \eq{Lquant}, 
we substitute the variables $\ell$ and $n_r$ by the classical actions $L$
and $S_r$, using $d\ell=dL/\hbar$ and $dn_r=dS_r/2\pi\hbar$. Then 
$g_{ebk}(e)$ becomes, expressed in terms of the dimensionless scaled variables,
\bea
g_{ebk}(e) = \frac{1}{2\pi\hbar^3}\,\frac{\omega^5}{\epsilon^2}\! 
             \sum_{N=-\infty}^\infty 
             \sum_{M=-\infty}^\infty\int_0^{l_m(e)} dl\,2l 
             \int_0^{s_r(e,l)} ds_r\,\delta(e-\widetilde{\!h}\,(s_r,l))\,
             e^{is\,[Ns_r+2\pi Ml]/\hbar-i\pi(N+M)} + \dots
\label{gebk2}
\eea
Hereby $s_r=S_r/s$ is the scaled radial action integral and 
$\,\widetilde{\!h}\,(s_r,l)$ the scaled Hamiltonian \eq{ham}. Note that the 
integration limits have been imposed by the energy conservation; $l_m(e)$ 
is the maximum scaled angular momentum at fixed energy. The lower limits
are, strictly speaking, $\hbar s/2$ for the integral over $l$ and $\hbar\pi 
s$ for the integral over $s_r$. Replacing them by zero corresponds to
neglecting corrections of higher order in $\hbar$. In the following, we
keep terms up to order $\hbar$ with respect to the leading factor $\propto
\hbar^{-3}$ in \eq{gebk2}. There are exactly two corrections of relative 
order $\hbar$ in what is indicated by the dots above: one is a boundary
correction from $n_r=0$ to \eq{gebk1}, and the second is coming from
the lower limit $\pi\hbar s$ of the $s_r$ integral in \eq{gebk2}. A short   
calculation shows that these two corrections cancel identically. All other 
terms neglected in \eq{gebk2} correspond to corrections of relative order 
$\hbar^2$ or higher.

We now use the relations
\bea
\delta (e-\widetilde{\!h}\,(s_r,l)) = |s_r'(e,l)|\,\delta(s_r-s_r(e,l))
\eea
and
\bea
s_r'(e,l) = \papa{s_r(e,l)}{e} = t_r(e,l)\,.
\eea
Here $t_r(e,l)=\omega T_r(e,l)$ is the scaled period of the classical motion 
at fixed values of $e$ and $l$, which is the energy derivative of the 
corresponding scaled action integral $s_r(e,l)$:
\bea
s_r(e,l) & = & \oint \tilde{p}_q(e,l)\,dq
        \; = \;2\int_{q_1}^{q_2}\! dq\sqrt{2e-2v(q)-l^2\!/q^2},
\label{Srad}\\
t_r(e,l) & = & \papa{s_r(e,l)}{e}
        \; = \;2\int_{q_1}^{q_2}\! dq\,
               \frac{1}{\sqrt{2e-2v(q)-l^2\!/q^2}}\,.
\label{Trad}
\eea
Here $q_1$ and $q_2$ are the scaled lower and upper turning points of the 
radial motion, respectively, which both depend on $e$ and $l$. The above 
integrals can in many cases be expressed in terms of complete elliptic 
integrals. (Exceptions are the HO and Coulomb potentials where they become 
simple algebraic functions of $e$ and $l$.) For our present potential $v(q)$ 
in \eq{escal}, the analytical expressions for \eq{Srad}, \eq{Trad} and other 
quantities of interest are given in the Appendix \ref{app3}. One important
result derived there is the Taylor expansion of $s_r(e,l)$ around $l=0$,
whose first terms are
\bea
s_r(e,l) = s_r(e,0) - \pi\,l + a(e)\,l^2 + {\cal O}(l^3)\,,
\label{SrLTay}
\eea
where $a(e)$ is given in \eq{aofe}. The structure of \eq{SrLTay} -- but not 
the explicit form of $a(e)$ -- appears to be a general result valid for all 
regular central potentials $v(q)$ with a minimum at $q=0$ (and hence not for 
the Coulomb potential, see Appendix \ref{appcoul}), but we were not able to 
prove it in the general case. The relevance of the linear term $-\pi\,l$ in 
\eq{SrLTay} will become clear in the next subsection.

Using the above relations we can now do the integral over $s_r$ in \eq{gebk2} 
exactly, due to the delta function, and obtain the following ``EBK trace 
integral'':
\bea
g_{ebk}(e) = \frac{1}{\pi\hbar^3}\,\frac{\omega^5}{\epsilon^2}\!
             \sum_{N=-\infty}^\infty 
             \sum_{M=-\infty}^\infty (-1)^{N+M}
             \int_0^{l_m(e)} dl\,l\,t_r(e,l)\,
             e^{i\,[NS_r(e,l)+2\pi Msl]/\hbar} + {\cal O}(\hbar^{-1})\,.
\label{gebkint}
\eea

The phase of the integrand of \eq{gebkint} can be interpreted as the full 
action $S_{N\!M}(e,l)=NS_r(e,l)+2\pi Msl$, divided by $\hbar$, of a given 
classical orbit labelled
by $M$ and $N$, consisting of the radial part $NS_r(e,l)$ and the angular part 
$2\pi Msl$. When doing the full double summation over all $M$ and $N$ and the 
integration over $l$ exactly, \eq{gebkint} yields the EBK spectrum \eq{eebk} 
to leading order in $\hbar$. Note that the limits of the $l$ integral are 
the cases of zero angular momentum, which corresponds to the diameter orbits, 
and its maximum value $l_m(e)$ at a given energy, which corresponds to the 
circle orbits. The latter have the radius $q_0$ at which the effective scaled 
potential $v_{eff}(q)$ has its minimum; for this motion the radial action is 
zero: $S_r(e,l_m(e))=0$, since all energy is in the angular motion. All 
contributions with $0 < l < l_m(e)$ to the integral correspond to motion 
which has both radial and angular components.

The formula \eq{gebkint}, valid for arbitrary (but correctly scaled) spherical 
potentials $v(q)$ with smooth walls, is 
in principle a trace formula, but it does not yet have its characteristic form. 
That form is obtained by evaluating the integral over $l$ in the semiclassical 
limit $\hbar\to 0$ by evaluating its leading contributions from the 
critical points of the phase function $S_{N\!M}(e,l)/\hbar$ in the exponent of 
the integrand. To leading order in $\hbar$, these are stationary points -- as 
far as they exist. Using the standard stationary-phase evaluation of the 
integral around the stationary points, one obtains contributions to $g_{ebk}(e)$ 
with amplitudes 
proportional to $\hbar^{-5/2}$, as expected for the fully three-fold degenerate 
orbits in a spherical system (cf.\ the discussion at the end of the previous 
section). Next to leading order in $\hbar$, one obtains contributions from the 
end points of the integral (see, e.g., \cite{wong}), sometimes referred to
as ``edge corrections''. As already announced above and shown explicitly below, 
these correspond here to the diameter and circular orbits, giving contributions 
with amplitudes proportional to $\hbar^{-2}$. 

A special contribution to $g_{ebk}(e)$ comes from $M=N=0$. As generally
proved by Berry and Mount \cite{bermo}, this must be the Thomas-Fermi
(TF) value of the density of states, which is the leading contribution
to its smooth part. From \eq{gebkint} we obtain with $M=N=0$
\bea
g_{ebk}^{(0)}(e) = \frac{1}{2\pi\hbar^3}\,\frac{\omega^5}{\epsilon^2}\!
                   \int_0^{l_m(e)}dl^2 t_r(e,l)\,.
\label{tfint}
\eea
The integral of \eq{Trad} over $l^2$ is straightforward. Note that the 
contribution from the upper limit $l_m(e)$ in \eq{tfint} is zero, since 
the turning points coincide: $q_1(e,l_m)=q_2(e,l_m)=q_0$. The result is
\bea
g_{ebk}^{(0)}(E) = \frac{\;2\omega^5}{\pi\hbar^3\epsilon^2} 
                   \int_0^{q_2^0}q^2dq\sqrt{2e-2v(q)}
                 = g_{TF}(E)\,.
\label{gtf}
\eea
Here $q_2^0$ is the upper turning point for $l=0$. (The lower one is zero: 
$q_1^0=0$.) That \eq{gtf} really is equal to the TF density of states 
follows from its general definition
\bea
g_{TF}(E) = \frac{1}{(2\pi\hbar)^3}\int\! d^3p \int\! d^3r\;
            \delta(E-\widetilde{H}(\bfr,\bfp))
          = \frac{s^3}{(2\pi\hbar)^3}\,\frac{\omega^5}{\epsilon^2}\!
            \int\! d^3\tilde{p} \int\! d^3q\;
            \delta(e-\widetilde{h}(\bfq,\bf\tilde{p}))\,.
\eea
Using polar coordinates for $\bfq$ and $\bf\tilde{p}$ and doing the 
$\bf\tilde{p}$ integration leads to \eq{gtf}. The analytical expression
valid for the potential \eq{escal} is given in \eq{gtf0}. In the 
introduction we have stated that the average part of the density of states 
is generally given by the ETF approximation which contains $\hbar$ 
corrections to the TF limit. For the spherical three-dimensional HO, these 
corrections are of relative order $\hbar^2$, as seen in \eq{getfho}. For 
the present perturbed HO potential \eq{hpert} we will see that the TF 
approximation is sufficient to reproduce the average part of the 
quantum-mechanical density of states, at least up to the energies for 
which the quantum spectrum is numerically available.

The stationary condition for the phase function in \eq{gebkint} at a point 
$l=l_{N\!M}$ reads
\bea
\papa{}{l}\,[Ns_r(e,l)+2\pi Ml]\Big|_{l_{N\!M}}
           = N\papa{s_r(e,l)}{l}\Big|_{l_{N\!M}}+2\pi M= 0\,.
\label{stat}
\eea
Due to energy conservation we may write
\bea
de = d\,\widetilde{\!h}\,(s_r,l) 
   = \papa{\widetilde{h}}{s_r}\Big|_l \,ds_r 
     + \papa{\widetilde{h}}{l}\Big|_{s_r}\! dl= 0\,,
\eea
so that 
\bea
\papa{s_r(e,l)}{l} = -2\pi\, \frac{\omega_\phi(e,l)}{\omega_r(e,l)}\,,
\label{dsrdl}
\eea
where the frequencies of angular and radial motion are defined as usual by
\bea
\omega_\phi(e,l) = \papa{\widetilde{H}(S_r,L)}{L}\,,\qquad
\omega_r(e,l) = 2\pi\,\papa{\widetilde{H}(S_r,L)}{S_r}\,.
\eea 
With this, the stationary condition \eq{stat} becomes
\bea
\frac{\omega_\phi(e,l_{N\!M})}{\omega_r(e,l_{N\!M})}=
\frac{T_r(e,l_{N\!M})}{T_\phi(e,l_{N\!M})} = \frac{M}{N}\,,
\label{tori}
\eea
which is the periodicity condition for the ``rational tori'' as the most 
degenerate classical orbits \cite{bertab}. Clearly, $N$ and $M$ must have 
the same sign, since frequencies and periods are positive quantities.
Because of \eq{SrLTay}, the lower integration limit $l=0$ in \eq{gebkint} is 
a stationary point corresponding to the diameter orbit with $N\!:\!M=2\!:\!1$. 
However, the stationary-phase integration of \eq{gebkint} for $l=l_{21}=0$ 
yields a zero contribution due to the factor $l$ in the integrand, so that 
the only semiclassical contribution due to the diameter orbits is the 
end-point correction for $l=0$ discussed in the next section.
As we shall see further below, solutions of \eq{tori} with $0 < l_{N\!M} 
< l_m(e)$ do not exist for all energies $e$ and for all pairs $N,M$ of 
integers. Therefore we postpone the contributions of the rational tori to 
\eq{gebkint} and concentrate first on the diameter and circle orbits. 

Before developing the corresponding trace formula for our present system, 
valid to all orders in $\epsilon$, we want to establish here the connection 
to the first-order perturbative approach used in Sect.\ \ref{secpert}. From 
the Taylor expansions given in Appendix \ref{app3}, we obtain for the radial 
action integral the following first terms:
\bea
2S_r(E,L) = 2\pi\left(\frac{E}{\omega}-L\right)
            -\frac{\epsilon\pi}{4\,\omega^5}\,(3E^2-\omega^2L^2)
            + {\cal O}(\epsilon^2)\,.
\label{Srad1}
\eea
In the second term we recognise precisely the first-order action shift
given in \eq{ds3d}. Inserting into it $L=0$ and the leading-order 
expression for $L_m=E/\omega$ given in \eq{Scirc}, we obtain the 
first-order action shifts of the diameter and circle orbits, respectively, 
given in \eq{sorbits}. We now insert \eq{Srad1} into the expression 
\eq{gebkint} for the density of states, neglecting the terms of higher 
order in $\epsilon$ and keeping only the zero-order terms in the amplitude 
of $T_r=\pi/\omega$, given in \eq{Trtay}, and of the upper integration 
limit $L_m=E/\omega$. Noting furthermore that for the unperturbed HO 
orbits the ratio $T_r/T_\phi$ becomes equal to 2, so that the resonance 
condition \eq{tori} implies $N=2M$, we assume for the moment that all other 
combinations of $N$ and $M$ in this limit may be neglected. Writing 
$N=2M=2k$, we obtain the following result for the density of states up to 
first order in $\epsilon$ (note that the linear terms in $L$ cancel in the 
exponent of the integrand):
\bea
g_{ebk}^{(1)}(E) = \frac{1}{2\hbar^3\omega}
                   \sum_{k=-\infty}^{\infty}(-1)^k\, e^{i2\pi kE/\hom} 
                   \int_0^{E/\omega}\! dL^2\,
                   e^{-ik\sigma(3-\omega^2\!L^2\!/E^2)]/\hbar},
\label{gebkp}
\eea
where $\sigma$ is the quantity defined in \eq{sigma}. Using the substitution 
$z=(\omega L/E)^2$, the above integral becomes identically equal to the 
integral for the perturbative modulation factor \eq{modfex}, and the result 
\eq{gebkp} is precisely the perturbed density of states defined in \eq{gpert}
with the oscillating part given in \eq{dgp}. 
Rather than proving at this stage that all contributions to \eq{gebkint} 
with $N\neq 2M$ either cancel or are negligible to the leading order in 
$\hbar$, we now go on to derive the full trace formula for the diameter and 
circle orbits, doing the summations over all $N$ and $M$ exactly in the 
semiclassical limit.

\subsection{Semiclassical trace formula for diameter and circle orbits
            in the perturbed harmonic-oscillator potential}
\label{sectrace}

Equipped with the above results, we are now in a position to evaluate the 
full trace formula for the contribution of diameter and circle orbits for 
our present Hamiltonian \eq{hpert}. The smooth TF part of the level 
density, $g_{TF}(E)$, is given explicitly in the appendix \ref{app3}. 
The contributions of the rational tori, which bifurcate 
from the circle orbits only for high enough energy and repetition numbers, 
will be discussed in Section \ref{sectori}. We therefore now evaluate the
end-point contributions to the integral in \eq{gebkint} in the
semiclassical limit $\hbar\to 0$.

\subsubsection{Diameter orbit}
\label{secdiam}

The lower end point $l=0$ in \eq{gebkint} yields the asymptotic contribution
\bea
\delta g_d(e) = \frac{\omega^3}{2\pi\epsilon\hbar^2}\,T_r(e,0)\!
                \sum_{N=-\infty}^\infty 
                (-1)^N e^{i[NS_r(e,0)/\hbar+\pi/2]}\,u_N(e)\,,
\label{gd1}
\eea
where we have left out the contribution from $N=0$ which contributes to the
TF part, and defined the quantity
\bea
u_N(e) = \lim_{l\to 0}\left[l\!\!\!\sum_{M=-\infty}^\infty 
         \frac{(-1)^M}{\left(\frac{N}{2}\papa{s_r}{l}+\pi M\right)}
         \right].
\eea
Using the identity \cite{abro}
\bea
\frac{1}{\sin(z)} = \sum_{M=-\infty}^\infty \frac{(-1)^M}{(z-M\pi)}\,,
                   \qquad\qquad (z\neq n\pi,\quad n\in\mathbb{Z})            
\label{polesum}
\eea
we find
\bea
u_N(e) = \lim_{l\to 0}\left[
         \frac{l}{\sin\left(\frac{N}{2}\papa{s_r}{l}\right)}\right].   
\label{une}
\eea
We now exploit the result \eq{SrLTay} from which we find
\bea
\papa{s_r}{l} = -\pi + 2a(e)\, l + {\cal O}(l^2)\,.
\eea
For odd values of $N$, the sin function in \eq{une} gives always a 
nonzero denominator in the limit $l=0$, so that $u_N(e)$ becomes zero. 
For even $N=2k$ we get in the limit $l\to 0$
\bea
\sin\left(\frac{N}{2}\papa{s_r}{l}\right) = \sin(-k\pi+2kla(e))
         = (-1)^k\sin(2kla(e))\;\longrightarrow\; (-1)^k 2kla(e)\,,
\eea
so that we obtain
\bea
u_{2k}(e) = \frac{(-1)^k}{2ka(e)}\,.
\eea
Inserting this into \eq{gd1} using \eq{STdia}, leaving out the contribution 
$k=0$ which is contained in $g_{TF}(e)$, we obtain the final contribution of 
the diameter orbits to the density of states:
\bea
\delta g_d(e) = -\frac{T_d(e)\,\omega^3}{4\pi\epsilon\hbar^2a(e)}\sum_{k=1}^\infty
                 \frac{(-1)^k}{k}\,\sin(kS_d(e)/\hbar)\,,
\label{dgd2}
\eea
where the actions $S_d(e)=2S_r(e,0)$ are given explicitly in \eq{Sr0}. 
In the low-energy limit we obtain with \eq{aofe} exactly the diameter 
contribution to the perturbative trace formula \eq{dgp}.

Note that the presence of the linear term $-\pi l$ in the quantity
\eq{SrLTay} is instrumental in annihilating the contributions with
odd $N$ to the sum in \eq{gd1} and hence in establishing the fact that
only an even number of radial oscillations yields a physical periodic
orbit with angular momentum $L=0$.

\subsubsection{Circle orbit}

The upper end point $l=l_m(e)$ in \eq{gebkint} yields the asymptotic 
contribution
\bea
g_c(e) = \frac{1}{\pi\hbar^2}L_m(e)\,T_r(e,l_m)\!
         \sum_{M=-\infty}^\infty (-1)^M
         e^{i[2\pi L_m(e)/\hbar-\pi/2]}
         \sum_{N=-\infty}^\infty \frac{(-1)^N}
         {\left(N\papa{s_r}{l}\big|_{l_m}\!\!+2\pi M\right)}\,.
\label{gc1}
\eea
Using \eq{dsrdl}, \eq{Scirc} and noting that $T_\phi(e,l_m)=T_c(e)$, 
we get 
\bea
\papa{s_r}{l}\Big|_{l_m} = -\frac{2\pi T_r(e,l_m)}{T_c(e)}\,.
\eea
Employing \eq{polesum} again to perform the $N$ summation, writing $|M|=k$ 
and omitting the $k=0$ term, we obtain after some manipulations the
contribution of the circle orbits to the density of states 
\bea
\delta g_c(e) = \frac{T_c(e)L_m(e)}{\pi\hbar^2}\sum_{k=1}^\infty
                \frac{(-1)^k}{\sin[k\pi w(e)]}\,\sin(kS_c(e)/\hbar)\,,
\label{dgc2}
\eea
where $S_c(e)$ is given explicitly in \eq{Scirc}, and the quantity $w(e)$
is defined by
\bea
w(e) = \frac{T_c(e)}{T_r(e,l_m)} = \frac{\omega_r(e,l_m)}{\omega_\phi(e,l_m)}.
\label{wofe}
\eea
In the limit $e\to 0$ this quantity is found with the r.h.s.\ of
the analytical results \eq{Tcirc} and \eq{trlm} to go like
\bea
w(e) = 2\,(1+e/4 + \dots) = 2+\frac{\epsilon}{2\omega^4}E + \dots
\eea
Expanding the denominator in \eq{dgc2}, using the above $w(e)$, up to first
order in $\epsilon$ brings the amplitude \eq{dgc2} exactly into that
of the circle orbit contribution to the perturbative result \eq{dgp}.

Note that the denominator under the sum in \eq{dgc2} looks exactly like that 
in the Gutzwiller trace formula \cite{gutz,mil} for an isolated stable orbit, 
whereby $\pi w(e)$ corresponds to one-half of the stability angle. Indeed, 
the semiclassical amplitude in \eq{dgc2} diverges when $kw(e)$ becomes an 
integer $n\in\mathbb{Z}$. At the corresponding nonzero energies, the circle 
orbit bifurcates; the condition for this to happen is exactly the resonance 
condition \eq{tori} for the rational tori with $M=k$, taken at 
$l_{N\!M}=l_m(e)$. (We come back to this point in Section \ref{sectori}.) 
Using \eq{Tcirc} and \eq{trlm}, we find that $w(e)$ is restricted to the 
following range:
\bea
2 < w(e) =  \frac{T_c(e)}{T_r(e,l_m)}
         < \sqrt{6} \qquad \hbox{for} \qquad 0 < e < \infty\,.
\label{wlim}
\eea
The smallest $k$ for which we can have $2<N/k<\sqrt{6}=2.4494897$ with 
$N\in{\mathbb N}$ is $k=3$ with $N=7$, so that $w(e_{7:3})=7/3$, which happens 
at the scaled energy $e_{7:3}\sim 7.670$ (see \tab{toribif} below). This means 
that the shortest new orbit is the 7:3 torus, bifurcating from the 3rd 
repetition of the circle orbit. The next bifurcation (from $k=4$) is that of 
the 9:4 torus at $e_{9:4}\sim 2.0967$. Still longer orbits bifurcate at lower 
energies, but from higher repetitions of the circle orbit. Therefore, we can 
ignore the bifurcations at low energies and small repetition numbers $k$ and 
concentrate on the contributions of the diameter and circle orbits to the 
density of states.

Let us finally write down the trace formula which we have obtained for
the oscillating part of the density of states:
\bea
\delta g(e) \simeq \sum_{k=1}^\infty \left[
            {\cal A}_k^c(e)\sin(kS_c(e)/\hbar)+
            {\cal A}_k^d(e)\sin(kS_d(e)/\hbar)\right]\!,
\label{dguni}
\eea
where the semiclassical amplitudes are given by
\bea
{\cal A}_k^c(e) = \frac{T_c(e)L_m(e)}{\pi\hbar^2}
                  \frac{(-1)^k}{\sin[k\pi w(e)]}\,,\qquad
{\cal A}_k^d(e) = \frac{T_d(e)\,\omega^3}{4\pi\epsilon\hbar^2a(e)}
                  \frac{(-1)^{k+1}}{k}\,.
\label{ampcd}
\eea
In the following we shall test this formula against the quantum-mechanical
density of states. As we have shown above, \eq{dguni} goes over into the 
perturbative trace formula \eq{dgp} in the limit $\epsilon\to 0$. 
This confirms the assumption, made at the end of Sect.\ \ref{secpois},
that the contributions with $N\neq 2M$ to \eq{gebkint} do not contribute 
in this limit.

\subsection{Numerical tests of the trace formulae}
\label{secqm}

In this section we compare our semiclassical results with those obtained 
from the exact quantum-mechanical spectrum (obtained numerically by solving 
the radial Schr\"odinger equation on a discrete mesh). In order to focus on 
the {\it coarse-grained} shell structure, we convolute both results over the 
energy $E$ with a normalised Gaussian of width $\gamma$. The 
quantum-mechanical density of states \eq{gqm} then becomes
\bea
g_{qm}(E) = \frac{1}{\gamma\sqrt{\pi}}\sum_n e^{-(E-E_n)^2/\gamma^2}.
\label{gqmgam}
\eea
In order to obtain its oscillating part, we subtract from it the TF
expression $g_{TF}(E)$ which we can calculate analytically (see Appendix
\ref{app3}). The semiclassical trace formula \eq{dgsc} becomes, after doing 
the convolution in stationary-phase approximation (cf.\ \cite{book}, Sec.\ 5.5)

\Figurebb{comp1}{230}{30}{795}{518}{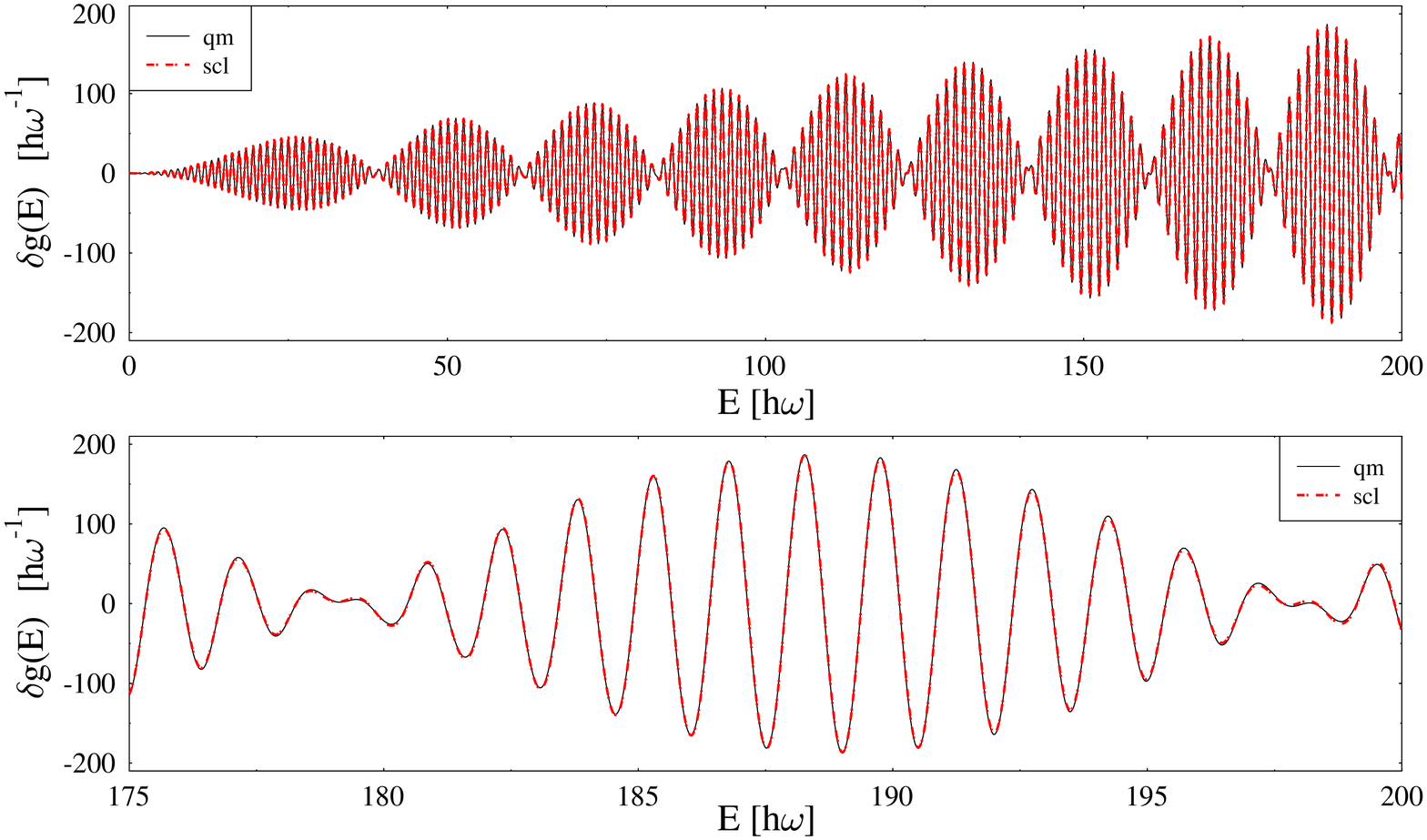}{9}{16.5}{
Density of states for the Hamiltonian \eq{hpert} with $\epsilon=0.01$, 
Gaussian-averaged with a width $\gamma=0.5\hom$, versus energy $E$ (units: 
$\hom$). {\it Dashed lines (red):} quantum-mechanical results.
{\it Solid lines (black):} semiclassical results using the uniform
trace formula \eq{dguni}.
}

\bea
\delta g_{sc}(E) \simeq \sum_{po} {\cal A}_{po}(E)\, 
                        e^{-[\gamma T_{po}(E)/2\hbar]^2}                        
                        \cos\,[S_{po}(E)/\hbar-\sigma_{po}\pi/2]\,,
\label{dgscgam}
\eea
so that orbits with longer periods $T_{po}$ will be exponentially
suppressed.

In \fig{comp1} we show a comparison of the results obtained with the
uniform trace formula \eq{dguni} for the case $\epsilon=0.01$ with
those obtained from the exact numerical quantum spectrum. The width
of the Gaussian smoothing was chosen to be $\gamma=0.5\hom$; in the 
semiclassical result the harmonics $k>2$ then do not contribute noticeably. 
The agreement is seen to be perfect; tiny differences can only be noted 
near the beat minima on the amplified scale below.

In \fig{comp2} we show a similar comparison for $\gamma=0.1\hom$, exhibiting
much fine structures. We now start to resolve the spectrum with a higher
resolution; note that we still only plot the oscillating part $\delta g(E)$
of the density of states. In the top panel, the semiclassical result includes 
harmonics (i.e., repetitions of the two orbits) with $k\leq k_m=7$. This is 
evidently sufficient to reproduce the quantum-mechanical oscillations up to  
$E\simeq 45\hom$ with a high accuracy. In the centre panel, we have added two 
more harmonics, going up to $k\leq k_m=9$. Here we recognise the appearance 
of divergences at the scaled energies $e=0.361712$ and $e=0.437867$ which 
correspond to the bifurcations of the 19:9 and 17:8 tori from the repetitions
of the circle orbit with $k=9$ and $k=8$, respectively (cf.\ \tab{toribif} 
below). The small wiggles close to the divergences, with decreasing amplitudes 
when going away from them, are due to the missing contributions from those 
torus orbits in the trace formula \eq{dguni} (see the following section).

\Figurebb{comp2}{75}{50}{1022}{745}{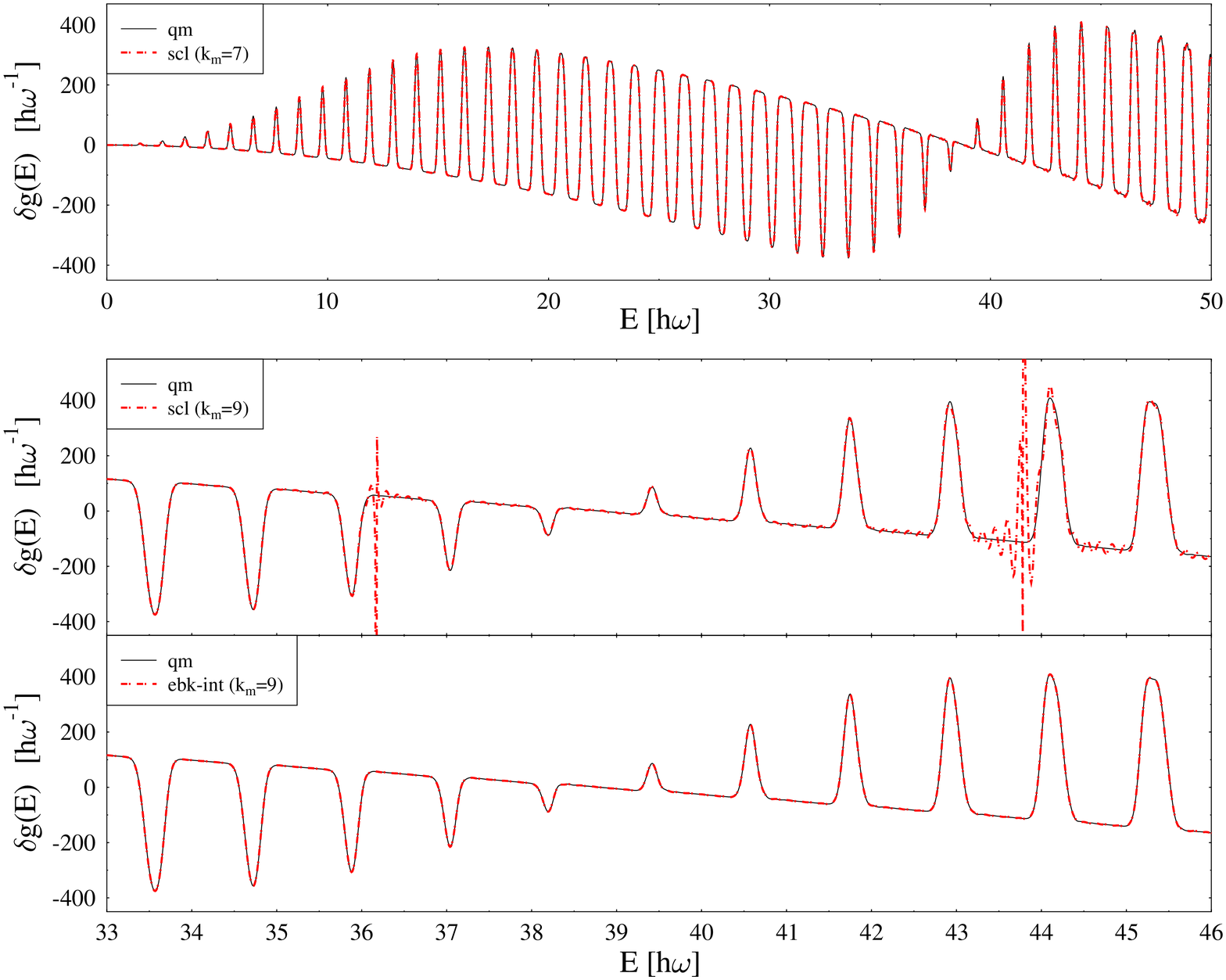}{13.5}{17}{
Same as \fig{comp1}, but with $\gamma=0.1\hom$. 
{\it Top panel:} including up to $k_m=7$ harmonics in the trace
formula \eq{dguni}.
{\it Centre panel:} adding harmonics up to $k_m=9$. Note the
divergences due to the bifurcations of the 17:8 and 19:9 tori at
the energies $E\simeq 43.79$ and $E\simeq 36.17$, respectively.
{\it Bottom panel:} result obtained from the EBK trace integral
\eq{gebkint}, doing the integration over $l$ numerically, with 
$-k_m\leq M\leq +k_m$, $-2k_m\leq N \leq +2k_m$ using $k_m=9$.
}

\vspace*{-0.2cm}

In order to anticipate the results of a suitable uniform trace formula
including the contributions of the tori, we show in the bottom panel of 
\fig{comp2} the result of the EBK trace integral \eq{gebkint} obtained by a 
numerical integration over the variable $l$. Hereby the summations over $N$ 
and $M$ have been done in the limits $-k_m\leq M\leq +k_m$, $-2k_m\leq N \leq 
+2k_m$ using $k_m=9$. [The period to be used in the Gaussian damping factor of 
\eq{dgscgam} is in this case given by $T_{po} = NT_r(e,L)$.] Now the 
divergences and all other discrepancies have disappeared; the difference
to the quantum-mechanical result can hardly be recognised. The integral
\eq{gebkint} therefore is also a uniform expression for the semiclassical
density of states, containing the contributions of all periodic orbits.
The numerical evaluation becomes, however, very time consuming for even
finer energy resolutions (i.e., still smaller values of $\gamma$).

Our main result here is that up to a rather fine resolution, the shell
structure of the present perturbed HO system is dominated by the two-fold
degenerate families of diameter and circle orbits, although these are not 
the most degenerate orbits. This is a rather unusual situation, due to the
fact that the shortest tori with the highest degeneracy are considerably
longer (by a factor $\simg$10) than the primitive diameter and circle
orbits. 

Let us finally test the range of validity of the perturbative trace
formula \eq{dgp}. In \fig{comppert} we compare its results to that
of the full trace formula \eq{dguni} for the two values $\epsilon=0.001$
(upper panel) and $\epsilon=0.01$ (lower panel). As expected, the two
curves agree in the limit $e\to 0$. However, already for
$\epsilon=0.001$ the amplitude modulation of the perturbative result
deviates so much that the position of the first beat node is predicted
by about 10\% too low in energy. For $\epsilon=0.01$, the perturbative
result becomes so bad that it predicts the second beat node approximately
at the position of the correct first one. These results underline the
fact that in \cite{yylund}, the quantum results for weak interactions
could be qualitatively reproduced by the perturbative trace formula, but
not quantitatively with correct positions of the beat nodes. A more
detailed analysis of the selfconsistent HF results of \cite{yylund} using 
the uniform trace formula \eq{dguni} is in progress \cite{magnus}.

\Figurebb{comppert}{230}{30}{795}{518}{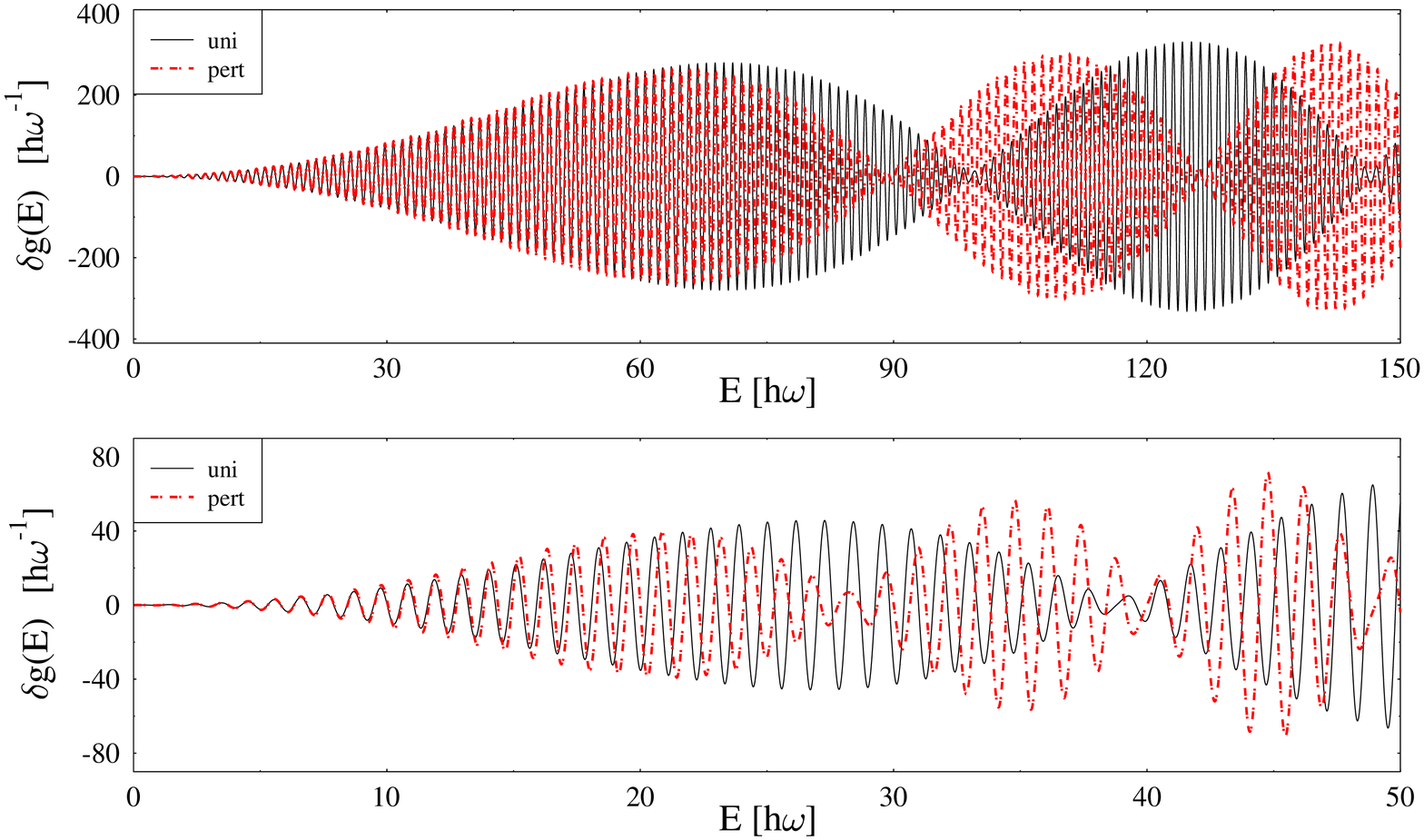}{9}{16.5}{
Same quantity as shown in \fig{comp1}. Here we compare the results of
the perturbative trace formula \eq{dgp} (dashed lines, red) and those
of the uniform trace formula \eq{dguni} (full lines, black), 
Gaussian-averaged with $\gamma=0.5\hom$. {\it Upper panel:} for
$\epsilon=0.001$; {\it lower panel:} for $\epsilon=0.01$.
}

\subsection{Bifurcations of rational tori from the circle orbit}
\label{sectori}

As we have seen in the previous section, periodic orbit with both angular and 
radial motion appear as the rational tori fulfilling the periodicity condition 
\eq{tori}. For the Hamiltonian \eq{hpert}, this condition can only be fulfilled 
in the limits given in \eq{wlim}. This means that stationary points $L_{N\!M}$ 
of the $L$ integral in \eq{gebkint} satisfying \eq{stat} and hence \eq{tori} 
only exist if $2<|N\!:\!M|<\sqrt{6}$. Hereby $|M|$ corresponds to the 
repetition number $k$ of the circle orbit that undergoes a bifurcation exactly 
when $L_{N\!M}$ is equal to the maximum angular momentum $L_m$. In \tab{toribif} 
we list some of the lowest energies $e_{N:M}$ at which this happens for the 
smallest integers $M=k$. Each of the $N\!\!:\!\!M$ tori only exists above its 
bifurcation energy, i.e., at energies $e>e_{N:M}$. 

\Table{toribif}{7.0}{
\begin{tabular}{|c|c||l|l||l|}
\hline
 $M=k$ & $N$ & $~~~e_{N:M}$ & $~l_{N\!M}$ & \;$l_{N\!M}^{(\epsilon=\infty)}$\\
\hline
\;\;\dots & \, \dots & \!\dots & \!\dots & \!\dots \\
9 & \,19\, & \,0.3617121\,& \,0.3370\; & \,0.276 \\
8 & \,17\, & \,0.4378671\,& \,0.4032\; & \,0.316 \\
7 & \,15\, & \,0.5527726\,& \,0.5009\; & \,0.360 \\
6 & \,13\, & \,0.7441150\,& \,0.6584\; & \,0.420 \\
5 & \,11\, & \,1.1174198\,& \,0.9508\; & \,0.504 \\
4 & \, 9\, & \,2.0966668\,& \,1.6553\; & \,0.633 \\
3 & \, 7\, & \,7.6699999\,& \,4.9331\; & \,0.864 \\
\hline
\end{tabular}\hspace*{0.5cm}
}{~~Bifurcation energies $e_{N:M}$ and angular momenta $l_{N\!M}$ of the 
rational $N\!\!:\!\!M$ tori at which the stationary conditions \eq{stat} and 
\eq{tori} are fulfilled for the lowest values of $M$ and $N$. In the
last column we give the angular momenta $l_{N\!M}$ for the limiting case
$\epsilon\to\infty$ in which they do not depend on the energy (cf.\
Sect.\ \ref{secquart}).
}

The contributions of the bifurcated tori to the semiclassical trace formula is 
obtained by a stationary-phase evaluation of the $l$ integral in \eq{gebkint}, 
leading to
\bea
\delta g_{tori}(e) = \sum_{N>2M>0} {\cal A}^T_{N\!M}(e)\,(-1)^{N\!+\!M}\!
                     \cos\left(\frac{S^T_{N\!M}(e)}{\hbar}+\frac{\pi}{4}\right).
\label{dgt}
\eea
The summation goes only over those pairs $N,M$ of positive integers for which 
the stationary condition \eq{tori} can be fulfilled. (Note that the contribution
of the corresponding pairs of negative integers is included in the amplitudes
by a factor 2.) The actions of the tori are given by
\bea
S^T_{N\!M}(e) = NS_r(e,l_{N\!M})+M2\pi L_{N\!M}(e)\,,
\label{stori}
\eea
and their amplitudes by
\bea
{\cal A}^T_{N\!M}(e) = \frac{2\sqrt{2s}}{\sqrt{\pi}\hbar^{5/2}}\,
                     T_r(e,l_{N\!M})L_{N\!M}(e)
                     \left(\!N\left|\frac{\partial^2s_r(e,l)}{\partial l^2}
                     \right|_{l=l_{N\!M}(e)}\right)^{\!\!-1/2}\theta(e-e_{N:M})\,,
\label{amptori}
\eea
where $s$ is the action unit given in \eq{sscal}. The Heaviside step function 
in \eq{amptori} is defined as
\bea
\theta(x) = 0 \quad\hbox{for}\quad x < 0\,,\qquad
\theta(x) = 1/2 \quad\hbox{for}\quad x = 0\,,\qquad
\theta(x) = 1 \quad\hbox{for}\quad x > 0\,,\qquad
\eea
where the value 1/2 for $x=0$ accounts for the fact that one obtains only 
half a Fresnel integral when the stationary point is the upper end point 
of the integral in \eq{gebkint}. The power $\hbar^{-5/2}$ in the amplitude 
${\cal A}_{N\!M}(e)$ is characteristic of the three-fold degenerate orbits 
in a three-dimensional spherical system \cite{struma,crli1}.

Eq.\ \eq{dgt} corresponds to the standard Berry-Tabor trace formula
\cite{bertab} for the most degenerate orbits in an integrable system.
Here, however, each contribution to \eq{dgt} is only valid if the energy $e$ 
is sufficiently larger than the corresponding bifurcation energy $e_{N:M}$.
In the neighbourhood of the bifurcation energies, we have to replace the
sum of the diverging circle orbit contribution and the corresponding
torus contribution with $|M|=r$ by a common uniform contribution. As
it is well-known from the semiclassical theory of bifurcations 
\cite{ozoha,ssun}, one has to include hereby also the contributions of
so-called ``ghost orbits'', which are the imaginary (or complex)
continuations of the bifurcated orbits (here: the tori) to the other
side of the bifurcation (here: $e<e_{N:M}$).

It is, however, not the purpose of our present paper to discuss in more 
detail the uniform approximation suitable for this situation. The present
scenario is actually almost identical to that described in \cite{kaidel}
for the bifurcations of tori in the integrable H\'enon-Heiles system.
Although there the orbit from which the tori bifurcate is an isolated
one, while it here is a family of circle orbits, the uniform trace formula 
given in \cite{kaidel} can be applied to the present system in a 
straightforward manner. We refer the interested reader to this paper for 
all the detailed formulae as well as for a numerical demonstration showing 
how diverging discrepancies of the type seen in the centre part of 
\fig{comp2} disappear when a proper uniform trace formula is used. 
Its result would here be practically indistinguishable from the result 
shown in the bottom part of \fig{comp2}, in which we have evaluated the
integral in \eq{gebkint} numerically instead of using the stationary phase
approximation with end-point corrections.

\subsection{The limit of the purely quartic oscillator}
\label{secquart}

In this section we discuss briefly a purely quartic oscillator potential,
i.e., we start from the Hamiltonian
\bea
H(\bfr,\bfp) = \frac12\,p^2 + \frac{\epsilon}{4}\,r^4 = E\,.
\label{hr4}
\eea
Here both the energy and $\epsilon$ can be scaled away. After dividing the 
above equation by $E$ and introducing the scaled variables 
\bea
q_i  = \left(\!\frac{\,\epsilon}{E}\right)^{\!1/4}\!r_i\,,\qquad 
\tau = (\epsilon E)^{1/4}\,t\,,\qquad
l    = \epsilon^{1/4}E^{-3/4}L\,,
\eea
we obtain the ``unit energy'' equation
\bea
1 = \frac12\,{\dot q}^2 + \frac14\,q^4 + \frac{l^2}{2q^2}\,,
\eea
where the dot again means derivative with respect to the scaled time $\tau$. 
Hence the classical equations of motion become independent of energy and 
$\epsilon$. After obtaining all the classical results, we can reintroduce 
$E$ and $\epsilon$ just by remembering that lengths scale with 
$(E/\epsilon)^{1/4}$, momenta with $\sqrt{E}$, times (periods) with 
$(E\epsilon)^{-1/4}$, actions and angular momenta with 
$\epsilon^{-1/4}E^{3/4}$.

We obtain the semiclassical trace formula from the results of the previous 
sections and of the appendix \ref{app3} simply by taking the limit 
$\epsilon\to\infty$, which means practically by extracting the 
leading terms for $e\to\infty$. In this way, we obtain the 
following primitive periods and actions for the circle orbit:
\bea 
T_c(E) = 2\pi\epsilon^{-1/4}(4E/3)^{-1/4}\,,\qquad
S_c(E) = 2\pi L_m(E) \; = \; 2\pi\epsilon^{-1/4}(4E/3)^{3/4}\,,
\label{circr4}
\eea
and for the diameter orbit:
\bea 
T_d(E) = 2\sqrt{2}\,\epsilon^{-1/4}K_0E^{-1/4}\,,\qquad
S_d(E) = 2\sqrt{2}\,\epsilon^{-1/4}(4/3)K_0E^{3/4}\,.
\label{diagr4}
\eea
Here $K_0$ is the constant elliptic integral
\bea
K_0=K(\kappa_0)=1.8540746773\dots\,,\qquad \kappa_0 = 1/\sqrt{2}\,.
\eea
Furthermore we get
\bea
T_r(E,L_m) = \pi(3\epsilon E)^{-1/4}\,.
\eea
The resonance condition \eq{tori} becomes independent of the energy
\bea
\frac{T_\phi(E,l_{N\!M})}{T_r(E,l_{N\!M})} = w(l_{N\!M}) = \frac{N}{M}\,,
\label{wofl}
\eea
since both periods scale with the same power of the energy. For the
frequency ratio we find the limits
\bea
2 \leq w(l_{N\!M}) = N\!:\!M < \sqrt{6}\qquad \hbox{for} \qquad 0 \leq l_{N\!M} 
                             < l_m = (4/3)^{3/4} = 1.240806\dots
\label{tori4}
\eea
Note that the upper limit $l_m$ never becomes a stationary point, so that no 
bifurcations of the circle orbits occur in this system. The values of the 
stationary points $0 < l_{N\!M} < l_m$ (in scaled dimensionless units) for the 
lowest values of $N$ and $M$ are given in the rightmost column of \tab{toribif}; 
as we just have seen, they do not depend on the energy $E$.

The result $w(l_m)=\sqrt{6}$ is a special case of the general result, given 
in \cite{arita}, that for a potential $V(r)$ = $U_0\,r^\beta$ one has 
$w(l_m)=\sqrt{\beta\!+\!2}$. 

For the amplitudes of the circle and diameter orbits in the trace formula 
\eq{dguni}, which keeps the same form, we obtain here 
\bea
{\cal A}_k^c(E) = \frac{4}{\hbar^2\sqrt{3}}\,\sqrt{\!\frac{E}{\epsilon}}\,
                  \frac{(-1)^k}{\sin(k\pi\sqrt{6})}\,,\qquad
{\cal A}_k^d(E) = \frac{4K_0^2}{\hbar^2\pi^2}\,\sqrt{\!\frac{E}{\epsilon}}\;
                  \frac{(-1)^{k+1}}{k}\,.
\label{ampr4}
\eea
The actions $S_c$ and $S_d$ to be used in \eq{dguni} are those given in
\eq{circr4} and \eq{diagr4}. Finally, the TF level density is found to be
\bea
g_{TF}(E) = \frac{4\sqrt{2}\,E^{5/4}}{5K_0\hbar^2\epsilon^{3/4}}\,.
\eea

The tori, which in the perturbed HO system discussed in the last section
occur only at sufficiently high energies after their bifurcations from the 
repeated circle orbits, exist in the present system at all energies. However, 
they affect again only the finer quantum structures of the density of states 
since they only exist for $N\!:\!M \geq 7\!:\!3$ due to the selection by their
resonance condition \eq{tori4} (cf.\ also \tab{toribif}). This is 
demonstrated in the following two figures.

In \fig{comp3}, we compare the results of the semiclassical trace formula
\eq{dguni} for the diameters and circles using the above quantities to the 
exact quantum-mechanical density of states, both Gaussian-averaged over a 
width $\gamma=0.5$ (in energy units). The agreement is again perfect even 
on the enlarged scale in the lower part of the figure. The reason is that
for the chosen energy resolution with $\gamma=0.5$, only repetition numbers 
$|M|=k<3$ contribute, for which there exist no tori (see \tab{toribif}).

\Figurebb{comp3}{70}{35}{994}{518}{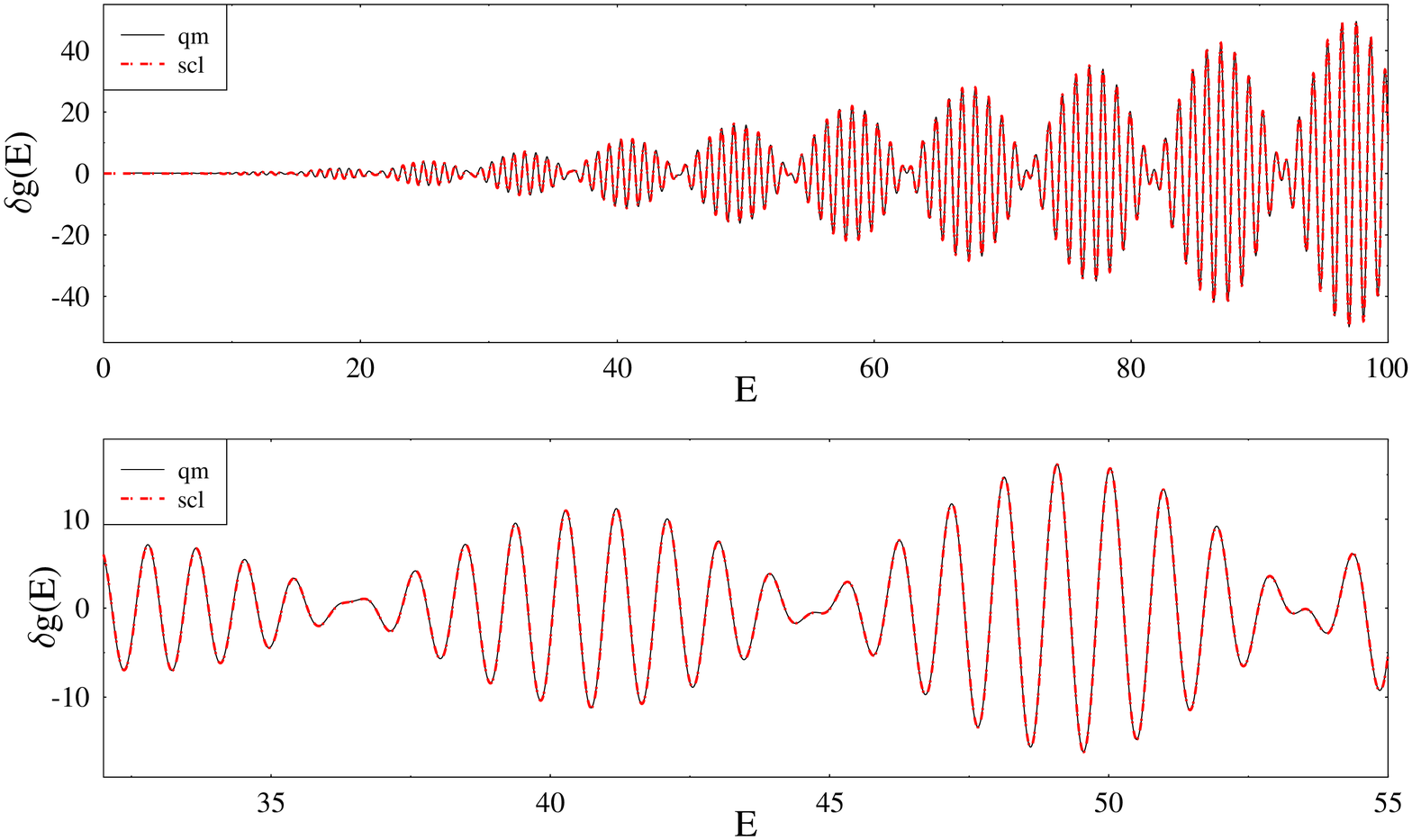}{9.2}{17}{
Density of states for the Hamiltonian \eq{hr4} with $\epsilon=0.01$, 
Gaussian-averaged with a width $\gamma=0.5$ energy units. 
{\it Dashed lines (red):} quantum-mechanical results.
{\it Solid lines (black):} semiclassical results using the uniform
trace formula \eq{dguni} and the actions and amplitudes given in
\eq{circr4}, \eq{diagr4} and \eq{ampr4} with $k\leq 2$.
}

\vspace*{-0.5cm}

\Figurebb{comp4}{80}{35}{994}{262}{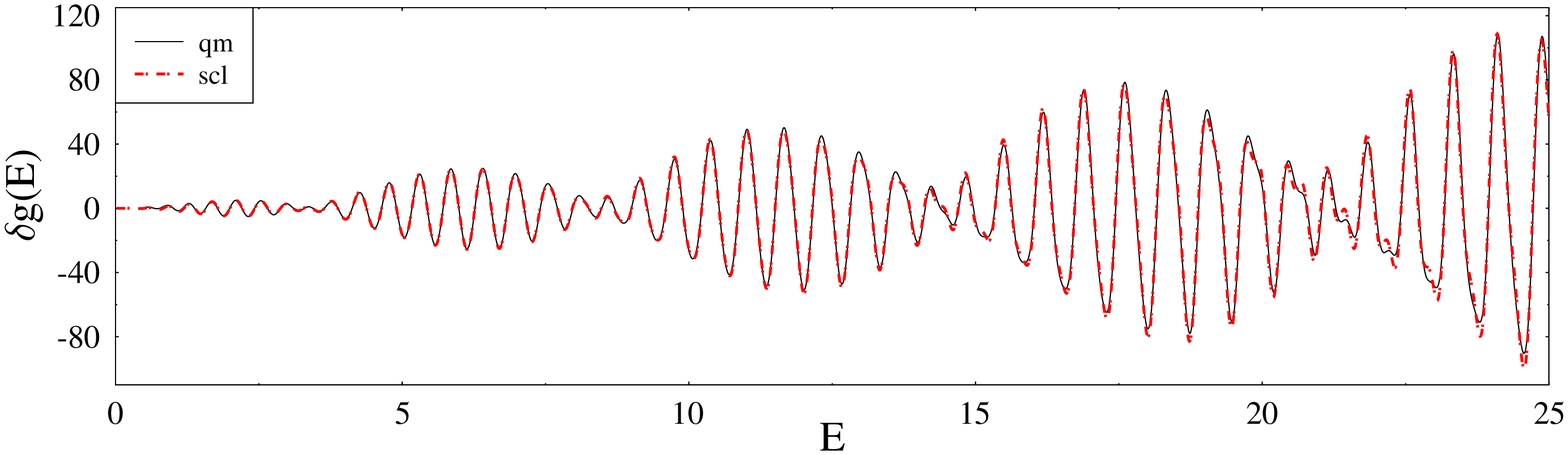}{4.5}{17}{
Same as \fig{comp3}, but with $\gamma=0.2$ energy units. Repetitions of
the diameter and circle orbits with $k\leq 5$ are included in the
semiclassical result.
}

\fig{comp4} shows the same comparison for $\gamma=0.2$. Here some small 
discrepancies appear which become more visible with increasing energy. They 
are due to the missing contributions from the tori with $|M|=k\geq 3$. 
Since there are no bifurcations involved in the present system, the 
contributions of the tori are given by the Berry-Tabor trace formula \eq{dgt}, 
employing the appropriate expressions for the actions and periods used 
therein. The analytical expression of the action integral $S_r(e,l)$ in terms 
of elliptic integrals is the same as for the perturbed HO potential, as given 
in Eqs.\ \eq{Srint} - \eq{kap}, except that here the term $q^2\!/2$ in the 
potential and the corresponding contribution $s_r^{(1)}(e,l)$ must be omitted.

Even at the rather fine resolution obtained in \fig{comp4}, our main result 
is the observation that the circle and diameter orbits dominate the shell 
structure also for the purely quartic oscillator. 

\section{Summary and conclusions}
\label{secsum}

Motivated by recent Hartree-Fock (HF) calculations of harmonically trapped
fermionic gases with a repulsive interaction \cite{yylund}, in which
pronounced super-shell effects were obtained, we have developed a 
semiclassical trace formula for the density of states of the 
anharmonically perturbed three-dimensional isotropic harmonic oscillator
(HO) defined in \eq{hpert}. We find that its gross-shell structure is 
dominated by the periodic orbits of diameter and circle shape, whose 
interference leads to super-shell beats which explain the findings from the 
numerical HF results.

In a first step, we have used the perturbative approach of Creagh \cite{crpert}
to describe the symmetry breaking U(3) $\to$ SO(3) for weak 
anharmonicities $\epsilon$, resulting in the perturbative trace formula \eq{dgp}. 
It uniformly restores the U(3) limit, yielding the exact trace formula 
\eq{g3hoc} of the unperturbed HO system in the limit $\epsilon\to 0$.
For the derivation of the perturbative result \eq{dgp}, one must in principle
integrate over the 4-fold degenerate families of unperturbed HO orbits
which cover the manifold $\mathbb{C}$P$^2$. As shown in App.\ \ref{app2},
however, it turns out to be easier to integrate over the energy shell in
phase space, which is a five-sphere S$^5$, whereby the integration can be
reduced, under exploitation of the SO(3) symmetry, to the one-dimensional integral
given analytically in \eq{modfex}. An interesting result hereby is the fact that 
its two contributions, corresponding to the diameter and circle orbits, already 
have the characteristic form occurring in the general trace formula \eq{dgsc}. 
Usually, this form is obtained from a perturbative trace formula only 
asymptotically in the semiclassical limit $\hbar\to 0$; this happens, 
e.g., also when one investigates the corresponding Hamiltonian \eq{hpert} in two 
dimensions or other perturbed 2-dimensional systems \cite{book,crpert,bcl}.

Next we have used the standard EBK quantisation and the Poisson summation 
formula to derive a general semiclassical trace formula for arbitrary central
potentials in terms of a one-dimensional integral over the angular momentum $L$
of the classical orbits, given in \eq{gebkint}. Exact integration and summation
over all $M,N$ yields, to leading order in $\hbar$, the EBK spectrum. Its
asymptotic evaluation in the limit $\hbar\to 0$ yields the standard-type
contributions to the trace formula \eq{dgsc}. The end-point contributions
yield the diameter orbits with $L=0$ and the circle orbits with maximum
angular momentum; both contributions are of next-to-leading order in $\hbar$,
whereas the leading-order contributions corresponding to the typical rational 
tori of integrable systems according to the Berry-Tabor theory \cite{bertab} 
come from the stationary points inside the integration interval. 
One interesting mathematical aspect is the mechanism by which 
the ratio $\omega_r:\omega_\phi=2$ of radial to angular frequency of the
diameter orbit is naturally selected by a property of the general radial 
action integral $S_r(E,L)=\oint p_r(E,L,r)\,dr$ as a function of angular 
momentum $L$ which appears to be universal for regular spherical potentials. 
It holds also for a particle in a spherical box with specular reflection, for 
which we have re-derived in Appendix \ref{appbb} the trace formula given by 
Balian and Bloch \cite{bablo} from our general formula \eq{gebkint}, taking 
into account the changes of the Maslov indices due to hard-wall reflections). 
The Coulomb potential, for which $\omega_r:\omega_\phi=1$, is discussed in 
Appendix \ref{appcoul} where we also re-derive the exact trace formula for the 
Rydberg spectrum given in \cite{book}.

For the Hamiltonian \eq{hpert}, we have derived a uniform trace formula
for the diameter and circle orbit contributions that is valid for arbitrary 
$\epsilon$. We find that for low energies and low repetition numbers 
no tori exist, so that the gross-shell structure of this 
system is dominated by the circle and diameter orbits and their lowest 
repetitions. The tori bifurcate, at sufficiently high energy, out of the 
repetitions with $k\geq 3$ of the circle orbits, as was also observed recently 
for homogeneous power-law potentials in \cite{arita}. Their contributions
sufficiently high above the respective bifurcation energies are given by
the usual Berry-Tabor type trace formula \eq{dgt}. We have not discussed here
the common uniform treatment of the circle orbits and the tori bifurcating
from them, since this scenario is identical to the one discussed in detail
in \cite{kaidel}.
The same qualitative results are found also in the limit $\epsilon\to
\infty$ in which the potential becomes a purely quartic oscillator. Although
the tori with $N\!:\!M\geq 7\!:\!3$ here exist at all energies, they are
so much longer (by a factor $\simg 10$) than the shortest diameter and circle 
orbits that they only affect the finer details of the quantum spectrum.

We thus have found the interesting and quite atypical situation of a
three-dimensional system in which the periodic orbits with highest, i.e., 
three-fold degeneracy are only responsible for finer details of the
quantum spectrum, whereas its gross-shell structure is to an astonishing
degree dominated by the orbits of next-to-leading order in $\hbar$, i.e.,
the diameter and circle orbits occurring in two-fold degenerate families. 
This is totally different from the situation, observed first by Balian and 
Bloch \cite{bablo}, of a spherical cavity potential (with ideally reflecting 
walls) which has turned out to be a realistic model for large spherical metal 
clusters. (We re-derive the semiclassical trace formula of Balian and Bloch 
for the spherical cavity in Appendix \ref{appbb}.) There the famous 
super-shells \cite{nish,klavs,mbclus} come from the interference of the 
shortest three-fold degenerate tori (the triangle and square orbits), whereas 
the two-fold degenerate diameter orbit family has virtually no effect on the 
observable shell structure of these systems. 

Our results explain qualitatively the numerical quantum-mechanical HF results  
\cite{yylund} for harmonically trapped fermionic atom gases. A more 
quantitative comparison, in which the strength $\epsilon$ of the perturbation 
in \eq{hpert} is determined directly from the numerically obtained 
self-consistent HF potentials, is in progress \cite{magnus}.

Lazzari {\it et al.}\ \cite{lazz} have studied the periodic orbits in a 
spherical Woods-Saxon potential in connection with the physics of metal
clusters and thereby focused on the contributions of the diameter orbits.
We find agreement of their results for the tori and the diameters with our 
results \eq{dgt}, \eq{amptori} and \eq{dgd2}, respectively, if we interpret 
their function $\tau(L_M,E)$ as being {\it half} the radial period 
$T_r(E,L_{N\!M})$ at the corresponding stationary value $L_{N\!M}$ of the 
angular momentum (= 0 for the diameters). They have, however, neglected 
the contributions from the circle orbits; furthermore they did not compare 
their semiclassical results with quantum-mechanical results. We suspect 
that the neglect of the circle orbits might be justifiable due to the chosen 
steepness of their Woods-Saxon potential, for which the triangle-like tori 
are perhaps more important than the circles.

Ozorio de Almeida {\it et al.}\ \cite{ozotom} have discussed the summation
of all periodic orbits in integrable systems in connection with spectral
determinants. They conclude that edge corrections corresponding to
lower-degenerate orbits may be neglected because they are implicitly
approximated by the nearest-lying rational tori (characterised by very 
large numbers $M,N$ in our notation). While their argument applies to the 
{\it full quantisation} of the exact (or EBK-approximated) quantum spectrum
obtained by summing over {\it all orbits}, it cannot be used when the 
spectrum is coarse-grained with a finite energy window $\gamma$, such as we 
have used the periodic-orbit sum in our present investigation (cf.\ also 
the discussion in the appendix of Ref.\ \cite{ozotom}) which is more
practically oriented.

\vspace*{0.1cm}

\acknowledgements

We acknowledge enlightening discussions with S. Bechtluft-Sachs, S. Creagh, 
and N. S{\o}ndergaard, and helpful comments by an anonymous referee. 
We are particularly grateful to K. J\"anich for 
providing us with the elegant reduction of the S$^5$ integration given in 
Appendix \ref{app2}. We thank A. Magner for valuable comments on our
manuscript and for bringing the Refs.\ \cite{lazz,ozotom} to our attention.
This work was financially supported by the Swedish 
Foundation for Strategic Research and the Swedish Research Council. 
M.\"O.\ acknowledges financial support from the Deutsche 
Forschungsgesellschaft (Graduiertenkolleg 638: {\it Nichtlinearit\"at und
Nichtgleichgewicht in kondensierter Materie}), and M.B.\ acknowledges the 
warm hospitality at the LTH during several research visits.

\begin{appendix}

\section{Parametrisation the manifold $\mathbb{C}$P$^2$}
\label{app1}

We reproduce here the parametrisation of $\mathbb{C}$P$^2$ given in \cite{bbz}.
We start with three complex numbers $Z_\alpha\in\mathbb{C}$:
\bea
(Z_1,Z_2,Z_3) = (n_1,n_2e^{i\nu_2},n_3e^{i\nu_3})\,,
\eea
with $n_\alpha \in \mathbb{R}$, $0\leq n_\alpha\leq 1\;(\alpha=1,2,3)$, and
$\nu_j\in \mathbb{R}$, $0\leq\nu_j<2\pi\; (j=2,3)$, 
whereby the $n_\alpha$ are restricted to
\bea
n_1^2+n_2^2+n_3^2=1\,.
\label{ensh}
\eea
The $Z_\alpha$ define the complex projective space $\mathbb{C}$P$^2$.
The $n_\alpha$ are in one-to-one correspondence with the points on
the first octant of the 2-sphere $S^2$, and the $\nu_j$ form a 2-torus.
At the edges of the octant, the torus contracts to a circle or to a
point, and one or both of the $\nu_j$ are not defined. 

Physically, we interpret the real and negative imaginary parts of the 
$Z_\alpha$ as coordinates and momenta, respectively, of a starting point
$(\bfr_0,\bfp_0)=(\{r_\alpha(0)\},\{p_\alpha(0)\})$ 
\bea
r_\alpha(0) = \Re e\, Z_\alpha\,,\qquad p_\alpha(0) = - \Im m\, Z_\alpha\,,
\eea
in the 6-dimensional phase space $(\bfr(t),\bfp(t))=(\{r_\alpha(t)\},
\{p_\alpha(t)\})$ of the 3-dimensional isotropic harmonic oscillator.
The Hamiltonian \eq{hho}, which has U(3) symmetry \cite{lipkin}, is given by
\bea
H(\bfr,\bfp) = \frac12 \sum_{\alpha=1}^3 \left[\omega^2r_\alpha^2+p_\alpha^2\right]
             = E\,.
\eea
The $2\pi$-periodic solutions of the equations of motion are then given by
\bea
r_\alpha(t) =  \Re e\, (Z_\alpha e^{i\omega t})\,,\qquad 
p_\alpha(t) = -\Im m\, (Z_\alpha e^{i\omega t})\,.
\eea
The equation \eq{ensh} defines the energy shell, so that the six quantities
$\{\omega r_\alpha,p_\alpha\}$ are points on a 5-sphere, S$^5$ (see also Appendix
\ref{app2} below). By taking the total 
energy to be $E=1/2$ and setting $\omega=1$, we fix the radius of the 
5-sphere to be unity. Taking out the phase factor $e^{i\omega t}$ takes us 
from S$^5$ to the projective space $\mathbb{C}$P$^2$.

Using the Fubini-Study metric, the authors of \cite{bbz} find the squared 
distance $ds^2$ between two points on $\mathbb{C}$P$^2$ to be
\bea
ds^2 = dn_1^2+dn_2^2+dn_3^2
     + n_2^2(1-n_2^2)\,d\nu_2^2+n_3^2(1-n_3^2)\,d\nu_3^2
     - 2\,n_2^2n_3^2\,d\nu_2d\nu_3\,.
\label{ds2}
\eea
The first three terms correspond to the standard metric on $S^2$, and
the last three to the metric on a flat 2-torus whose shape depends on
where we are on the octant.

We now wish to re-parametrise the $n_\alpha$ in terms of the two polar angles
$(\vartheta,\varphi)$ on $S^2$ by writing
\bea
n_1 = \cos\vartheta\,,\qquad
n_2 = \sin\vartheta\cos\varphi\,,\qquad
n_3 = \sin\vartheta\sin\varphi\,,
\label{coord}
\eea
with $\vartheta,\varphi\in[0,\pi/2]$. Collecting our four angles in a
vector $\bgrk{\psi}$ by defining 
\bea
\bgrk{\psi} = (\psi_1,\psi_2,\psi_3,\psi_4) = (\vartheta,\varphi,\nu_2,\nu_3)\,,
\eea
we can write the squared distance $ds^2$ as
\bea
ds^2 = \sum_{i,j=1}^4 g_{ij}\,d\psi_i\,d\psi_j\,,
\label{ds2a}
\eea
and find the elements $g_{ij}$ of the metric tensor to be
$g_{11} = 1$,   
$g_{22} = \sin^2\!\vartheta$,
$g_{12} = g_{21} = 0$, 
$g_{33} = \sin^2\!\vartheta\cos^2\!\varphi\,(1-\sin^2\!\vartheta\cos^2\!\varphi)$,
$g_{44} = \sin^2\!\vartheta\sin^2\!\varphi\,(1-\sin^2\!\vartheta\sin^2\!\varphi)$,
and $g_{34} = g_{43} = - \sin^4\!\vartheta\cos^2\!\varphi\sin^2\!\varphi$.
From this, we get the integration measure on $\mathbb{C}$P$^2$ to be
\bea
d\Omega_{\mathbb{C}{\rm P}^2} = |{\rm det}(g_{ij})|^{1/2}
                           d\psi_1d\psi_2\,d\psi_3\,d\psi_4
                     = \sin^3\!\vartheta\cos\vartheta\,d\vartheta\,
                           \cos\varphi\sin\varphi\,
                           d\varphi\;d\nu_2d\nu_3\,.  
\eea
The integrated volume of $\mathbb{C}$P$^2$, consistent with a general 
formula for $\mathbb{C}$P$^n$, is
\bea
\Omega_{\mathbb{C}{\rm P}^2} = \int d\Omega_{\mathbb{C}{\rm P}^2} 
                              = \frac{\pi^2}{2\,}
     \quad \Leftarrow \quad \Omega_{\mathbb{C}{\rm P}^n} = \frac{\pi^n}{n!}\,.
\eea

\section{Integration of the modulation factor over S$^5$}
\label{app2}

Here we give a strict proof of Eq.\ \eq{modfex} of the modulation factor
${\cal M}$ for the anharmonically perturbed HO system \eq{hpert}. The most 
straightforward parametrisation of the six-dimensional phase space of 
the unperturbed HO is given in terms of the coordinate vector 
$\bfr(t)=\{r_\alpha(t)\}$ and the momentum vector $\bfp(t)=\{p_\alpha(t)\}$
$(\alpha=1,2,3)$. Since a periodic solution is uniquely determined by their 
initial values at $t=0$: $\bfr_0=\bfr(0)$ and $\bfp_0=\bfp(0)$, and 
the total energy $E=(\bfp^2+\omega^2\bfr^2)/2$ is a constant of motion, the 
six components of $\bfr_0$ and $\bfp_0/\omega$ must cover a 
5-sphere with radius $R=\sqrt{2E}/\omega$. We introduce dimensionless 
coordinate and momentum vectors $\bgrk{\rho}$ and $\bgrk{\pi}$, respectively, 
by
\bea
\bgrk{\rho} = \frac{1}{R}\,\bfr_0\,,\qquad
\bgrk{\pi}  = \frac{1}{R\omega}\,\bfp_0\,,
\label{unvec}
\eea
so that in these variables the energy shell becomes the unit sphere S$^5$:
\bea
\bgrk{\rho}^2+\bgrk{\pi}^2 = 1\,.
\label{ens5}
\eea
From Eq.\ \eq{ds3d}, the first-order action perturbation $\Delta S_1$ 
is given in terms of energy $E$ and angular momentum $L$ by
\bea
\Delta S_1 = -\sigma\,(3-\omega^2L^2\!/E^2)\,,
\label{ds1}
\eea
with $\sigma$ defined in \eq{sigma}. The modulation factor for the first-order 
perturbed HO system becomes therefore
\bea
{\cal M}(k\sigma/\hbar) = \frac{1}{\pi^3} \!\int d\Omega_{S^5} \,
                          e^{ik\Delta S_1\!/\hbar},
\label{modfac5}
\eea
where $\Omega_{S^5}$ is the five-dimensional solid angle and $d\Omega_{S^5}$
is the integration measure on S$^5$ with $\int d\Omega_{S^5}=\pi^3$. One
may now express $\Delta S_1$ in \eq{ds1} directly in terms of the five
polar angles of six-dimensional hyperspherical coordinates, which becomes a 
very complicated function, and integrate \eq{modfac5} over all five angles. 
We have done this numerically and verified that it yields exactly the same 
result as the numerical 4-dimensional $\mathbb{C}$P$^2$ integration of
\eq{mfac} with the phase function \eq{ds3d}.

It is, however, not necessary to perform the full five-dimensional integral for
\eq{modfac5}. 
We follow a much more economical route \cite{jaen}, exploiting the rotational 
symmetry of the system and the SO(3) invariance of the expression \eq{ds1} for 
$\Delta S_1$. Due to the restriction \eq{ens5}, we can write the S$^5$ sphere as
\bea
S^5 = \left\{(\cos\vartheta\,{\bf e}_{\rho},\sin\vartheta\,{\bf e}_{\pi})
      \Bigr|\; \vartheta \in [0,\pi/2]\,;\; 
      {\bf e}_{\rho},{\bf e}_{\pi}\in S^2\right\},
\label{s5new}
\eea
where ${\bf e}_{\rho}$ and ${\bf e}_{\pi}$ are the unit vectors in the
directions of $\bgrk{\rho}$ and $\bgrk{\pi}$, respectively. The square
of the conserved total angular momentum then is
\bea
L^2 = {\bf L}^2 = (\bfr_0\times\bfp_0)^2 
                = \frac{4E^2}{\omega^2}\cos^2\!\vartheta\,\sin^2\!\vartheta
                  \,({\bf e}_{\rho}\times{\bf e}_{\pi})^2\,,
\eea
so that the action perturbation becomes, after the substitution $\alpha = 
2\vartheta$,
\bea
\Delta S_1 = -\sigma[\,3-\sin^2\!\alpha\,({\bf e}_{\rho}\times{\bf e}_{\pi})^2]\,,
             \qquad \alpha \in [0,\pi].
\label{ds1a}
\eea
The integration measure for \eq{s5new} is 
\bea
d\Omega_{S^5} = \cos^2\!\vartheta\,d\Omega_{S^2}({\bf e}_{\rho})\, 
                \sin^2\!\vartheta\,d\Omega_{S^2}({\bf e}_{\pi})\,d\vartheta
              = \frac18\sin^2\!\alpha\,d\alpha\,
                d\Omega_{S^2}({\bf e}_{\rho})\, d\Omega_{S^2}({\bf e}_{\pi})\,,
\eea
and the modulation factor becomes
\bea
{\cal M}(k\sigma/\hbar) = \frac{1}{8\pi^3} \!\int_0^{\pi}d\alpha\,\sin^2\!\alpha
                          \int d\Omega_{S^2}({\bf e}_{\rho})
                          \int d\Omega_{S^2}({\bf e}_{\pi})\,
                          e^{-ik\sigma[\,3-\sin^2\!\alpha\,({\bf e}_{\rho}
                                                 \times{\bf e}_{\pi})^2]/\hbar}.
\label{mf5a}
\eea
We now introduce the angle $\theta$ between the unit vectors ${\bf e}_{\rho}$ 
and ${\bf e}_{\pi}$, so that
\bea
({\bf e}_{\rho}\times{\bf e}_{\pi})^2 = \sin^2\theta\,,
\eea
and choose ${\bf e}_{\rho}$ as the direction of the north pole ($\theta=0$)
of polar angles $(\theta,\phi)$ for the vector ${\bf e}_{\pi}$. The
integrand then becomes independent of ${\bf e}_{\rho}$, so that the
corresponding S$^2$ integral just gives $\int\!d\Omega_{S^2}({\bf e}_{\rho})=4\pi$.
For the other S$^2$ integral we have $\int\!d\Omega_{S^2}({\bf e}_{\pi})=2\pi\!
\int_0^\pi\!\sin\theta\,d\theta$ since the integrand does not depend on $\phi$. 
We thus obtain, after the standard substitution $u=\cos\theta$ and an obvious 
reduction of the integration limits,
\bea
{\cal M}(k\sigma/\hbar) = \frac{4}{\pi}\int_0^{\pi/2} d\alpha\,\sin^2\!\alpha
                          \int_0^1 du\, 
                          e^{-ik\sigma[\,3-\sin^2\!\alpha\,(1-u^2)]/\hbar}.
\eea 
Next, we make the substitution $t=u\sin\alpha$ and then a further
substitution $s=\cos\alpha$ to obtain
\bea
{\cal M}(k\sigma/\hbar) = \frac{4}{\pi}\int_0^{\pi/2}d\alpha\,\sin\alpha
                          \int_0^{\sin\alpha} dt\,
                          e^{-ik\sigma(3-\sin^2\!\alpha+t^2)/\hbar} 
                        = \frac{4}{\pi}\int_0^1 ds
                          \int_0^{\sqrt{1-s^2}} dt\, 
                          e^{-ik\sigma(2+s^2+t^2)/\hbar}.
\eea 
Since the last integral goes exactly over the first quadrant of a unit 
disk in the $(s,t)$ plane, we can use polar coordinates $(r,\varphi)$ and 
furthermore $r^2=1-z$ to obtain our final result
\bea
{\cal M}(k\sigma/\hbar) = \frac{4}{\pi}\int_0^{\pi/2} d\varphi
                          \int_0^1 dr\,r\,e^{-ik\sigma(2+r^2)/\hbar}
                        = \int_0^1 dz\,e^{-ik\sigma(3-z)/\hbar},
\eea 
which is identical to that given in \eq{modfex} or in \eq{gebkp}.

\section{Explicit integrals for the anharmonically perturbed HO}
\label{app3}

In this appendix we derive analytical expressions for the particular scaled 
potential $v(q)$ in \eq{escal}. Rewriting the 
radial action integral \eq{Srad} using the substitution $q^2=z$, we obtain
\bea
s_r(e,l) = 2\int_{q_1}^{q_2}\! dq\sqrt{2e-q^2-q^4/2-l^2\!/q^2}
         = \frac{1}{\sqrt{2}}\int_{z_1}^{z_2}\frac{dz}{z}\sqrt{g(z)}\,,
\label{Srint}
\eea
where
\bea
g(z) = 4ez-2z^2-z^3-2l^2 = (z_1-z)(z_2-z)(z_3-z)\,.
\label{gz}
\eea
The classical turning points are defined by the roots of the cubic
equation $g(z)=0$, which are always real: the discriminant can easily
be shown to be negative except for $l=l_m(e)$ where it becomes zero due 
to the double root $z_1=z_2$. Selecting the roots such that
\bea
z_3 < z_1 \leq z_2\,,
\eea
one finds that $z_3$ is always negative, whereas $z_1$ and $z_2$ are
positive definite and represent the squares of the real classical turning
points $q_1$ and $q_2$. 

The integrals in \eq{Srint} cannot be found in most tables. However,
after an integration by parts we may write the r.h.s.\ as
\bea
s_r(e,l) = s_r^{(1)}(e,l)+ s_r^{(2)}(e,l)-l^2 s_r^{(-1)}(e,l)\,, 
\label{srltot}
\eea
where we have defined
\bea
s_r^{(n)}(e,l) = \sqrt{2}\int_{z_1}^{z_2} dz \frac{z^n}{\sqrt{g(z)}}\,. 
\eea
In Byrd and Friedman \cite{byfr}, these integrals are found to be:
\bea
s_r^{(1)}(e,l) & = &  \frac{2\sqrt{2}}{\sqrt{z_2-z_3}}\frac{z_2}{\kappa^2}\,
                      [(\kappa^2-\alpha^2)\,K(\kappa)+\alpha^2E\,(\kappa)]\,,
                      \nonumber\\
s_r^{(2)}(e,l) & = &  \frac{2\sqrt{2}}{\sqrt{z_2-z_3}}\frac{z_2^2}{3\kappa^4}\,
                      [(3\kappa^4-6\alpha^2\kappa^2+2\alpha^4+\kappa^2\alpha^4)\,
                      K(\kappa)+2\alpha^2(3\kappa^2-\alpha^2-\kappa^2\alpha^2)\,
                      E(\kappa)]\,,\nonumber\\
s_r^{(-1)}(e,l) & = & \frac{2\sqrt{2}}{\sqrt{z_2-z_3}}\frac{1}{z_2}\,
                      \Pi(\alpha^2,\kappa)\,.
\label{sradn}
\eea
Here $K(\kappa)$, $E(\kappa)$ and $\Pi(\alpha^2,\kappa)$ are the complete
elliptic integrals of first, second and third kind with modulus $\kappa$,
and
\bea
\kappa   = \sqrt{\frac{z_2-z_1}{z_2-z_3}}\,,\qquad 
\alpha^2 = 1-\frac{z_1}{z_2}\,.
\label{kap}
\eea 
Note that $0\leq\kappa^2\leq\alpha^2\leq 1$, which corresponds to the 'circular 
case' of the elliptic integral $\Pi(\alpha^2,\kappa)$ \cite{byfr}. This function
becomes singular like $(1-\alpha^2)^{-1/2}$, i.e., for $\alpha\to 1$ which 
happens when $l\to 0$. Since $z_1$ goes to zero like $l^2\!/2e$ in this
limit, $\Pi(\alpha^2,\kappa)$ has a first-order pole at $l=0$. Using the Laurent 
expansion of $\Pi(\alpha^2,\kappa)$ given in Eq.\ 906.04 of \cite{byfr}, we find
\bea
s_r^{(-1)}(e,l) = \frac{\sqrt{2}\pi}{\sqrt{-z_1z_2z_3}} -2a(e) + {\cal O}(l)
                = \frac{\pi}{l} -2a(e) + {\cal O}(l) \,,
\label{srlm1}
\eea
with
\bea
a(e) = - \frac{\sqrt{2}}{z_2^0\sqrt{z_2^0-z_3^0}}
         \left[K(\kappa_0)-\frac{E(\kappa_0)}{(1-\kappa_0^2)}\right]
     = \frac{\pi}{8}\left(1-\frac{15}{8}\,e +\dots\right).
\label{aofe}
\eea
Here $z_i^0$ and $\kappa_0$ are the $l=0$ values of the quantities given in 
Eqs.\ \eq{zi0} and \eq{kap0} below. The last equality in \eq{srlm1} follows 
because $z_1z_2z_3=-2l^2$, as seen from \eq{gz}. The functions $s_r^{(n)}(e,l)$ 
with $n=1,2$ are both regular in $l=0$. Altogether we obtain the following 
Taylor expansion for the total radial action 
integral \eq{srltot}:
\bea
s_r(e,l) = s_r(e,0) - \pi\,l + a(e)\,l^2 + {\cal O}(l^3)\,,
\label{SrLexp}
\eea

The radial period \eq{Trad} is easily found to be
\bea
t_r(e,l) = s_r^{(0)}(e,l) = \frac{\;2\sqrt{2}}{\sqrt{z_2-z_3}}\,K(\kappa)\,.
\label{Trad1}
\eea

In the limit $l=0$, which is relevant for the diameter orbits and the TF 
density of states, the above results simplify considerably. We obtain
\bea
s_r(e,0)  & = & \frac23\,(1+4e)^{1/4}\left[\left(\sqrt{(1+4e)}+1\right)
                K(\kappa_0)-2E(\kappa_0)\right], 
\label{Sr0}\\
t_r(e,0)  & = & 2\,(1+4e)^{-1/4}K(\kappa_0)
\label{Tr0}\,.
\eea       
Note that in this limit we get
\bea
z_1^0=0\,,\qquad z_2^0=(q_2^0)^2=\sqrt{1+4e}-1\,,\qquad z_3^0=-\sqrt{1+4e}-1\,,
\label{zi0}
\eea
and the modulus of the elliptic integrals becomes
\bea
\kappa_0 = \sqrt{\frac{z_2^0-z_1^0}{z_2^0-z_3^0}}
         = \sqrt{\frac{\sqrt{1+4e}-1}{2\sqrt{1+4e}}}
         = \sqrt{e}\left(1-\frac{3}{2}\,e +\dots\right).
\label{kap0}
\eea
The TF expression \eq{gtf} for the density of states becomes
\bea
g_{TF}(e) & = & \frac{8}{15\pi\hbar^3}\,\frac{\omega^5}{\epsilon^2}
                \,(1+4e)^{1/4}\,
                [2(1+3e)E(\kappa_0)-(1+3e+\sqrt{1+4e})K(\kappa_0)]\,.
\label{gtf0}
\eea
It is of interest to see the first terms in the Taylor expansion of the 
above results around $e=0$, whose leading terms are those of the pure
harmonic oscillator:
\bea
s_r(e,l)  & = & \pi\,(e-l)-\frac18\,\pi\,(3e^2-l^2)+\dots\,,
                \label{Srtay}\\
t_r(e,l)  & = & \pi -\frac34\,\pi\,e+\frac{15}{64}\,\pi(7e^2-l^2)+\dots\,,
                \label{Trtay}\\
g_{TF}(e) & = & \frac{1}{2\hbar^3}\,\frac{\omega^5}{\epsilon^2}\,
                e^2(1 - 5e/4 + \dots)\,.\label{gtftay}
\eea
The unscaled leading HO terms are
\bea
S_r^{ho}(E,L) = \pi\!\left(\frac{E}{\omega}-L\right) = \pi[L_m(E)-L]\,,\qquad
T_r^{ho} = \frac{\pi}{\omega}\,,\qquad
g_{TF}^{ho}(E) = \frac{E^2}{2(\hom)^3}\,. 
\eea

The diameter orbit with $L=0$ makes two full radial oscillations during its
primitive period. Therefore its primitive action and period are given by
\bea
S_d(e) = 2S_r(e,0)\,,\qquad T_d(e)=2T_r(e,0)\,.
\label{STdia}
\eea
For the circle orbit, the primitive action integral $S_c(e)$ is just $2\pi$ 
times the maximum value of the angular momentum, $L_m(e)=s\,l_m(e)$, which 
for the present potential is found to be
\bea
L_m(e) = s\sqrt{\frac{8}{27}\,[(1+3e)^{3/2}-1-9e/2]}\,,\qquad
S_c(e) = 2\pi L_m(e) = \frac{2\pi E}{\omega}\,(1-e/4+\dots)\,.
\label{Scirc}
\eea
The primitive period of the circle orbit is given by its energy derivative 
\bea
T_c(e) = \pi\frac{\sqrt{6}}{\omega}\,
         \frac{(1+3e)^{1/2}-1}{\sqrt{(1+3e)^{3/2}-1-9e/2}}
       = \frac{2\pi}{\omega}\,(1-e/2+\dots)\,.
\label{Tcirc}
\eea
We also give here the value of the radial period $T_r(e,l)$ at $l=l_m(e)$.
The corresponding frequency $\omega_r(e,l_m)=2\pi/T_r(e,l_m)$ is easily 
obtained from the second derivative of the effective potential at its 
minimum $q_0$, which for the present potential is given by
\bea
q_0^2=z_0=2(\sqrt{1+3e}-1)/3\,,
\eea
using $\omega_r^2(e,l_m)=V''_{eff}(q_0)$, and the result becomes
\bea
T_r(e,l_m) = \frac{2\pi}{\omega_r(e,l_m)}
           = \frac{\pi}{\omega}\,(1+3e)^{-1/4} 
           = \frac{\pi}{\omega}\,(1-3e/4+\dots)\,.
\label{trlm}
\eea
This agrees with the result obtained from \eq{Trad1}, noting that for the 
maximum value of $l=l_m(e)$, $z_1=z_2=z_0$ and $z_3=-2(2\sqrt{1+3e}+1)/3$, 
so that $\kappa$ \eq{kap} becomes zero and $K(0)=\pi/2$.

\section{Trace formulae for other spherical potentials}
\label{appothers}

In order to illustrate the validity and the usage of our general EBK trace 
integral \eq{gebkint}, we present in this appendix briefly its application 
to two popular spherical potentials.

\vspace*{-0.2cm}

\subsection{The spherical box potential}
\label{appbb}

We consider a particle with mass $m=1$ in a spherical box with radius $R$ 
and ideal specular reflection from the boundary, i.e., a three-dimensional
spherical billiard. Quantum-mechanically, one obtains the spectrum by solving 
the Schr\"odinger equation with Dirichlet boundary conditions. The 
semiclassical trace formula for this system has been derived by Balian and 
Bloch \cite{bablo} using a multi-reflection expansion of the Green function. 
To derive it from our formula \eq{gebkint}, we need the radial action integral 
for arbitrary energy $E\geq 0$ and angular momentum $0\leq L\leq L_m=R\sqrt{2E}$. 
Working throughout with unscaled variables, we find (with $r_1=L/\sqrt{2E}$)
\bea
S_r(E,L) = 2\int_{r_1}^R dr\sqrt{2E-L^2\!/r^2}
       & = & 2\sqrt{2ER^2-L^2} - 2L\arccos\,(L/R\sqrt{2E}) \nonumber\\
       & = & 2R\sqrt{2E} - \pi L + L^2\!/R\sqrt{2E} + {\cal O}(L^3)\,,
\label{SrLbox}
\eea
thus fulfilling our general relation \eq{SrLexp} with 
\bea
a(e) = 1/R\sqrt{2E}\,.
\eea
With this, the stationary condition \eq{tori} gives
\bea
2N\arccos\,(L/R\sqrt{2E}) = 2\pi M
\eea
with the solutions
\bea
L_{N\!M} = R\sqrt{2E}\cos(\pi M/N). 
\eea
As in the potentials discussed in Sect.\ \ref{secuni}, the solution with
$L=0$ corresponding to N=2M contributes only via the end-point correction
of \eq{gebkint} to the diameter orbit with the primitive action
\bea
S_d(E) = 2S_r(E,0) = 4R\sqrt{2E}
\eea
and, according to the lower equation in \eq{ampcd}, with the amplitude
of its $k$-th repetition
\bea
{\cal A}_k^d(E) = \frac{R^2}{\pi\hbar^2}\frac{1}{k}\,. \qquad (k=|M|)
\eea
The upper end-point correction with $L_{N\!M}=L_m=R\sqrt{2E}$ can, for $M\neq 0$, 
only be reached formally in the limit $N\to\infty$. It corresponds 
to the ``whispering gallery mode'' with amplitude ${\cal A}\propto 1/\sqrt{N}$,
cf.\ \eq{ampbb} below, and does therefore not contribute to the trace formula. 
The tori with $0<L_{N\!M}<L_m$ have the actions
\bea
 S^T_{N\!M}(E) = NS_r(E,L_{N\!M})+2\pi M L_{N\!M}
               = 2NR\sqrt{2E}\sin\varphi_{N\!M}\,, \qquad 
\varphi_{N\!M} = \pi M/N\,.
\label{stbb}
\eea
This corresponds exactly to the actions of the polygonal orbits given in
\cite{bablo} with winding number $k=|M|\geq 1$ and $v=|N|>2k$ corners (i.e., 
reflections from the boundary). The case $v=2k$ corresponds to the $k$-th 
repetition of the diameter orbit. For the polygons, $\varphi_{vk}=\pi k/v$ is 
half the polar angle covered by one of their segments. The amplitudes of the 
tori with $v>2k$ become, according to \eq{amptori},
\bea
{\cal A}^T_{vk}(E) = \frac{2R^{5/2}(2E)^{1/4}}{\hbar^{5/2}\sqrt{\pi}}
                     \sin(2\varphi_{vk})\sqrt{\frac{\sin\varphi_{vk}}{v}}\,.
\label{ampbb}
\eea

Before we can write down the trace formula with the correct phases, we have 
to correct the radial quantisation condition \eq{Irad}, because the Maslov 
index changes from 1/4 to 1/2 per turning point if a reflection happens there.
Hence the quantisation condition is $S_r=2\pi\hbar(n_r+1)$ for the diameter
orbit and $S_r=2\pi\hbar(n_r+3/4)$ for the tori. This changes the phases in
\eq{gebkint} and all the results derived from it. Taking this into account, 
we obtain the semiclassical trace formula for the spherical billiard
\bea
\delta g(E) = \sum_{v>2k>0}\!\!{\cal A}^T_{vk}(E)\,
              \sin[S^T_{vk}(E)/\hbar\!-\!3v\pi\!/2\!-\!k\pi+3\pi\!/4]
            - \sum_{k>0} {\cal A}_k^d(E)\,\sin[kS_d(E)/\hbar]\,,
\label{dgbb}                         
\eea
which is identically the result of Balian and Bloch \cite{bablo}.

\subsection{The Coulomb potential}
\label{appcoul}

Here we derive the trace formula for the Coulomb potential as a special 
central potential which is not regular in $r=0$:
\bea
V(r) = -\frac{\alpha}{r}\,.
\label{vcoul}
\eea
We need not use scaled variables here, but it is useful to introduce
the positive energy $e=-E$ since all bound states in the potential 
\eq{vcoul} have $E<0$. The maximum angular momentum and the action of
the circle orbit are
\bea
L_m(e) = \frac{\alpha}{\sqrt{2e}}\,,\qquad S_c(e)=\frac{2\pi\alpha}{\sqrt{2e}}\,,
\eea
from which the period of the circle orbit is found to be
\bea
T_c(e) = 2\pi \frac{dL_m(e)}{dE} = \frac{\pi\alpha}{\sqrt{2}\,e^{3/2}}\,.
\eea
The radial action integral becomes elementary (cf.\ \cite{book,ajp}):
\bea
S_r(e,L) = \sqrt{2}\int_{r_1}^{r_2} dr \sqrt{-e+\alpha/r-L^2\!/2r^2}
         = 2\pi\!\left(\frac{\alpha}{\sqrt{2e}}-L\right)
         = 2\pi[L_m(e)-L]\,.
\label{Srcoul}
\eea
From it we find the radial period
\bea
T_r(e) = \frac{\pi\alpha}{\sqrt{2}\,e^{3/2}}\,.
\eea
The difference from the other potentials treated in this paper
is that the term linear in $L$ in \eq{Srcoul} here has the coefficient 
$-2\pi$. Furthermore, we find from the general relation \eq{dsrdl}
\bea
\papa{S_r(e,L)}{L} = -2\pi \qquad \Rightarrow \qquad
    \omega_\phi(e) = \omega_r(e)\,, \qquad T_r(e) = T_\phi(e)\,.
\label{Trat}
\eea
Thus, the radial and angular frequencies are identically the same for 
all values of the angular momentum $L$. This is a special property of 
Kepler's ellipses. [Note that here, in contrast to the spherical HO, the
angular momentum $L$ is defined with respect to one of the focal points
and not to the symmetry centre of the ellipse!]

Inserting the above results into the general EBK trace integral 
\eq{gebkint} for the density of states, we obtain
\bea
g_{coul}(e) = \frac{T_r(e)}{\pi\hbar^3}\! \sum_{M,N=-\infty}^\infty\!
              (-1)^{M+N}\int_0^{L_m(e)}\! LdL\,
              e^{i2\pi[N(L_m-L)+ML]/\hbar}.
\label{gcoul}
\eea
Due to \eq{Trat}, the general resonance condition \eq{tori} implies $N=M$ 
for all periodic orbits. We therefore only use the terms with $N=M=\pm k$ in
the double sum of \eq{gcoul} (cf.\ the discussion at the end of Sect.\ 
\ref{secpois}). We then find
\bea
g_{coul}(e) = \frac{\alpha^3}{\hbar^3(2e)^{5/2}}
              \left[1+2\sum_{k=1}^\infty \cos\left(\frac{k}{\hbar}
              \frac{2\pi\alpha}{\sqrt{2e}}\right)\!\right].
\eea
Writing this in terms of atomic units (note that we have put $m=1$)
\bea
E_{at}=\frac{\alpha^2}{2\hbar^2} = Ry\,,
\eea
we finally obtain the quantum-mechanically exact trace formula
\bea
g_{coul}(E)  = \sum_{n=1}^\infty n^2 \delta(E+E_{at}/n^2)
             = \frac{E_{at}^{3/2}}{2(-E)^{5/2}}
               \left[1+2\sum_{k=1}^\infty \cos\left(2\pi k
               \sqrt{-E_{at}/E}\right)\!\right]\!,
\label{gcoul1}
\eea
which has been given in \cite{book}. The eigenenergies
\bea
E_n = -\frac{E_{at}}{n^2}\,,\qquad \qquad (n=n_r+\ell+1)
\eea
with degeneracies $d_n=n^2$ are, as is well known, also found directly from 
the EBK quantisation conditions \eq{Irad} - \eq{Iphi} using \eq{Srcoul} and 
\eq{Lquant}.

\end{appendix}


\newpage

\begin{thebibliography}{99}

\bibitem{gutz}   M. C. Gutzwiller, J. Math.\ Phys.\ {\bf 12}, 343 (1971),
                 and earlier papers quoted therein.

\bibitem{bablo}  R. Balian and C. Bloch, Ann.\ Phys.\ (N.Y.)
                 {\bf 69}, 76 (1972).

\bibitem{struma} V. M. Strutinsky, Nukleonika (Poland) {\bf 20}, 679 (1975);\\
                 V. M. Strutinsky and A. G. Magner, Sov.\ J. Part.\ Nucl.\ 
                 {\bf 7}, 138 (1976) [Elem.\ Part.\ \& Nucl.\ (Atomizdat, 
                 Moscow) {\bf 7}, 356 (1976)].

\bibitem{crli1}  S. C. Creagh and R. G. Littlejohn, Phys.\ Rev.\ A {\bf 44}, 
                 836 (1991); S. C. Creagh, J. Phys.\ A {\bf 26}, 95 (1993).

\bibitem{crli2}  S. C. Creagh and R. G. Littlejohn, J. Phys.\ A {\bf 25}, 
                 1643 (1992).

\bibitem{book}   M. Brack and R. K. Bhaduri: {\it Semiclassical Physics} 
                 (revised edition, Westview Press, Boulder, USA, 2003).

\bibitem{chaos}  M. C. Gutzwiller: {\it Chaos in Classical and Quantum
                 Mechanics} (Springer Verlag, New York, 1990);\\
                 H.-J. St\"ockmann: {\it Quantum Chaos: an Introduction}
                 (Cambridge University Press, Cambridge, UK, 1999);\\
                 F. Haake: {\it Quantum Signatures of Chaos} (Springer,
                 2nd edition, 2001).

\bibitem{bertab} M. V. Berry and M. Tabor, Proc.\ R. Soc.\ Lond.\ A {\bf 349},
                 101 (1976).

\bibitem{ebk}    A. Einstein, Verh.\ Dtsch.\ Phys.\ Ges.\ {\bf 19}, 82 (1917);\\
                 L. Brillouin, J. Phys.\ Radium {\bf 7}, 353 (1926);\\
                 J. B. Keller, Ann.\ Phys.\ (N. Y.) {\bf 4}, 180 (1958).

\bibitem{bm}     Aa. Bohr and B. R. Mottelson: {\it Nuclear Structure}
                 Vol.\ I (World Scientific, New York, 1975).

\bibitem{nish}   H. Nishioka, K. Hansen, and B. R. Mottelson, 
                 Phys.\ Rev.\ B {\bf 42}, 9377 (1990).

\bibitem{klavs}  J. Pedersen, S. Bj{\o}rnholm, J. Borggren, K. Hansen,
                 T. P. Martin, and H. D. Rasmussen, Nature {\bf 353}, 733 (1991).

\bibitem{mbclus} M. Brack, Rev.\ Mod.\ Phys.\ {\bf 65}, 677 (1993); 
                 The Scientific American, December 1997, p.\ 50.

\bibitem{steffi} S. M. Reimann and M. Manninen,
                 Rev.\ Mod.\ Phys.\ {\bf 74}, 1283 (2002).

\bibitem{wire}   A. I. Yanson, I. K. Yanson, and J. M. van Ruitenbeek,
                 Nature {\bf 400}, 144 (1999); Phys.\ Rev.\ Lett.\ 
                 {\bf 84}, 5832 (2000).

\bibitem{ozoha}  A. M. Ozorio de Almeida and J. H. Hannay, J. Phys.\ A {\bf 20},
                 5873 (1987).

\bibitem{toms}   S. Tomsovic, M. Grinberg, and D. Ullmo, Phys.\ Rev.\ Lett.\ 
                 {\bf 75}, 4346 (1995).
             
\bibitem{hhuni}  M. Brack, P. Meier, and K. Tanaka, J. Phys.\ A {\bf 32}, 331 
                 (1999).

\bibitem{ssun}   M. Sieber, J. Phys.\ A {\bf 29}, 4715 (1996);\\
                 H. Schomerus and M. Sieber, J. Phys.\ A {\bf 30}, 4537 (1997);\\ 
                 M. Sieber and H. Schomerus, J. Phys.\ A {\bf 31}, 165 (1998).

\bibitem{lipkin} see, e.g., H. J. Lipkin: {\it Lie Groups for Pedestrians} 
                 (North-Holland, Amsterdam, 1965).

\bibitem{yylund} Y. Yu, M. \"{O}gren, S. \AA berg, S. M. Reimann, and M. Brack,
                 preprint arXiv:cond-mat/0502096 (2005).

\bibitem{crpert} S. C. Creagh, Ann.\ Phys.\ (N. Y.) {\bf 248}, 60 (1996).

\bibitem{bbz}    I. Bengtsson, J. Br\"annlund, and K. \.{Z}yczkowski, 
                 Int.\ J. Mod.\ Phys.\ A {\bf 17}, 4675 (2002).

\bibitem{ajp}    see also L. J. Curtis and D. G. Ellis, Am.\ J. Phys.\ {\bf 72}, 
                 1521 (2004), for a pedagogical and concise presentation of EBK
                 quantisation of spherical systems.

\bibitem{titch}  see, e.g., E. C. Titchmarsh: {\it Introduction to the
                 theory of Fourier Integrals} (Second edition, Clarendon
                 Press, Oxford, 1948), p.\ 60.

\bibitem{wong}   R. Wong: {\it Asymptotic Approximation of Integrals}
                 (Academic Press Inc., San Diego, 1989).

\bibitem{bermo}  M. V. Berry and K. E. Mount, Rep.\ Prog.\ Phys.\ {\bf 35},
                 315 (1972).

\bibitem{abro}   M. Abramowitz and I. A. Stegun: {\it Handbook 
                 of Mathematical Functions}, 9th printing (Dover, New York, 1970)

\bibitem{mil}    For stable orbits, see also W. H. Miller, J. Chem.\ Phys.\ 
                 {\bf 63}, 996 (1975).

\bibitem{magnus} M. \"Ogren, work in progress.  

\bibitem{kaidel} J. Kaidel and M. Brack, Phys.\ Rev.\ E {\bf 70}, 016206 (2004).

\bibitem{arita}  K. Arita, Int.\ J. Mod.\ Phys.\ E {\bf 13}, 191 (2004).

\bibitem{bcl}    M. Brack, S. C. Creagh, and J. Law,
                 Phys.\ Rev.\ A {\bf 57}, 788 (1998).

\bibitem{lazz}   G. Lazzari, H. Nishioka, E. Vigezzi, and R. Broglia,
                 Phys.\ Rev.\ B {\bf 53}, 1064 (1996).

\bibitem{ozotom} A. M. Ozorio de Almeida, C. H. Lewenkopf, and S. Tomsovic,
                 J. Phys.\ A {\bf 35}, 10629 (2002). 

\bibitem{jaen}   K. J\"anich, private communication (2005).

\bibitem{byfr}   P. F. Byrd and M. D. Friedman: {\it Handbook of Elliptic
                 Integrals for Engineers and Scientists} (Springer-Verlag, 
                 Berlin, 2nd revised edition, 1971).

\end{thebibliography}
\end{document}